\documentclass[longauth]{aa} % for the long lists of affiliations 
\usepackage{graphicx}
\usepackage{lscape}
\usepackage{txfonts}
\usepackage{array,longtable}

\begin{document}

   \title{A luminous stellar outburst during a long-lasting eruptive phase first, and then SN IIn 2018cnf}
   \titlerunning{An outburst followed by SN IIn 2018cnf}

   \author{A. Pastorello\inst{1}
          \and
          A. Reguitti\inst{1}
          \and
          A. Morales-Garoffolo\inst{2}
          \and
           Z. Cano\inst{3,4}
          \and
          S.~J. Prentice\inst{5}
          \and
          D. Hiramatsu\inst{6,7}
          \and
          J.~Burke\inst{6,7}
          \and
          E.~Kankare\inst{5,8}
          \and
          R. Kotak\inst{8}
          \and
          T.~Reynolds\inst{8}
          \and
          S.~J. Smartt\inst{5}
          \and
          S. Bose\inst{9}
          \and
          Ping Chen\inst{9}
          \and
          E. Congiu\inst{10,11,12}
          \and
          Subo Dong\inst{9}
          \and
          S.~Geier\inst{13,14}
          \and
          M.~Gromadzki\inst{15}
          \and
          E.~Y.~Hsiao\inst{16}
          \and
          S.~Kumar\inst{16}
          \and
          P.~Ochner\inst{1,10}
          \and
          G.~Pignata\inst{17,18}
          \and
          L.~Tomasella\inst{1}
          \and
          L.~Wang\inst{19,20}
          \and
          I.~Arcavi\inst{21}
          \and
          C.~Ashall\inst{16}
          \and
          E.~Callis\inst{22}
          \and
          A.~de~Ugarte Postigo\inst{3,23}
          \and
          M.~Fraser\inst{22}
          \and
          G.~Hosseinzadeh\inst{24}
          \and
          D.~A.~Howell\inst{6,7}
          \and          
          C.~Inserra\inst{25}
           \and
          D.~A.~Kann\inst{3}  
          \and
          E.~Mason\inst{26}
          \and
          P.~A.~Mazzali\inst{27,28}
          \and
          C.~McCully\inst{7}
          \and
          \'O.~Rodr\'iguez\inst{17,18}
          \and
          M.~M.~Phillips\inst{12}
          \and
          K.~W.~Smith\inst{5} 
          \and
          L.~Tartaglia\inst{29}
          \and
          C.~C.~Th\"one\inst{3}
          \and
          T.~Wevers\inst{30}
          \and
          D.~R.~Young\inst{5}
          \and
          M.~L.~Pumo\inst{31}
          \and
          T.~B.~Lowe\inst{32}  
          \and
          E.~A.~Magnier\inst{32} 
          \and
          R.~J.~Wainscoat\inst{32} 
          \and
          C. Waters\inst{32} 
          \and
          D. E. Wright\inst{33} }

   \institute{INAF - Osservatorio Astronomico di Padova, Vicolo dell'Osservatorio 5, I-35122 Padova, Italy
              \email{andrea.pastorello@inaf.it}
         \and
    Department of Applied Physics, University of C\'adiz, Campus of Puerto Real, E-11510 C\'adiz, Spain 
         \and
     Instituto de Astrof\'isica de Andaluc\'ia (IAA-CSIC), Glorieta de la Astronom\'ia s/n, E-18008, Granada, Spain         
         \and
    Berkshire College of Agriculture, Hall Place, Burchetts Green Rd, Burchett's Green, Maidenhead, UK
         \and
     Astrophysics Research Centre, School of Mathematics and Physics, Queen's University Belfast, Belfast BT7 1NN, UK
         \and
    Department of Physics, University of California, Santa Barbara, CA 93106-9530, USA
         \and
   Las Cumbres Observatory, 6740 Cortona Drive, Suite 102, Goleta, CA 93117-5575, USA
         \and
   Department of Physics and Astronomy, University of Turku, FI-20014 Turku, Finland
         \and
   Kavli Institute for Astronomy and Astrophysics, Peking University, Yi He Yuan Road 5, Hai Dian District, Beijing 100871, China
         \and
   Dipartimento di Fisica e Astronomia, Universit\`a di Padova, Vicolo dell'Osservatorio 3, I-35122 Padova, Italy
         \and
   INAF - Osservatorio Astronomico di Brera, via E. Bianchi 46, 23807 Merate (LC), Italy
         \and
   Las Campanas Observatory - Carnagie Institution of Washington, Colina el Pino, Casilla 601, La Serena, Chile
         \and
   Gran Telescopio Canarias (GRANTECAN), Cuesta de San Jos\'e s/n, E-38712, Bre\~na Baja, La Palma, Spain
         \and
   Instituto de Astrof\'isica de Canarias, V\'ia L\'actea s/n, E-38200, La Laguna, Tenerife, Spain
         \and
   Warsaw University Astronomical Observatory, Al. Ujazdowskie 4, 00-478 Warszawa, Poland        
         \and
   Department of Physics, Florida State University, Tallahassee, FL 32306, USA
         \and
  Departamento de Ciencias F\'isicas, Universidad Andr\'es Bello, Santiago, Chile
         \and
   Millennium Institute of Astrophysics, Santiago, Chile
         \and  
  National Astronomical Observatories, Chinese Academy of Sciences, Beijing 100101, China
         \and
   Chinese Academy of Sciences South America Center for Astronomy, China-Chile Joint Center for Astronomy, Camino El Observatorio 1515, Las Condes, Santiago, Chile
         \and 
  The School of Physics and Astronomy, Tel Aviv University, Tel Aviv 69978, Israel
         \and  
  School of Physics, O'Brien Centre for Science North, University College Dublin, Belfield Dublin 4, Ireland
      \and
   Dark Cosmology Centre, Niels Bohr Institute, Juliane Maries Vej 30, Copenhagen Ø, 2100, Denmark
         \and 
  Center for Astrophysics | Harvard $\&$ Smithsonian, 60 Garden Street, Cambridge, MA 02138-1516 USA
         \and  
  School of Physics $\&$ Astronomy, Cardiff University, Queens Buildings, The Parade, Cardiff CF24 3AA, UK
      \and
  INAF - Osservatorio Astronomico di Trieste, Via G.B. Tiepolo 11, I-34143 Trieste, Italy
        \and  
  Astrophysics Research Institute, Liverpool John Moores University, 146 Brownlow Hill, Liverpool L3 5RF, UK
        \and
  Max-Planck Institut f\"ur Astrophysik, Karl-Schwarzschild-Str. 1, 85748 Garching, Germany
      \and
  The Oskar Klein Centre, Department of Astronomy, Stockholm University, AlbaNova, 10691 Stockholm, Sweden
        \and  
  Institute of Astronomy, Madingley Road, Cambridge CB3 0HA, United Kingdom
      \and
  Dipartimento di Fisica e Astronomia, Universit\`a degli Studi di Catania, Via S. Sofia 64, I-95123 Catania, Italy
      \and
  Institute for Astronomy, University of Hawaii, 2680 Woodlawn Drive, Honolulu, HI 96822, USA
      \and
  School of Physics and Astronomy, University of Minnesota, 116 Church Street SE, Minneapolis, Minnesota 55455-0149, USA
      }
   \date{Received 6 March 2019 / Accepted 8 July 2019}

  \abstract
   {We present the results of the monitoring campaign of the 
Type IIn supernova (SN) 2018cnf (a.k.a. ASASSN-18mr). It was discovered about ten days before the maximum light
(on MJD=58\,293.4 $\pm$ 5.7 in the $V$ band, with $M_V=-18.13\pm0.15$ mag). 
The multiband light curves show an immediate post-peak decline with some minor luminosity fluctuations, followed 
by a flattening starting about 40 days after maximum. The early spectra are relatively blue and 
show narrow Balmer lines with P Cygni profiles. Additionally, Fe II, O I, He I, and Ca II are 
detected. The spectra show little evolution with time and with intermediate-width features becoming progressively 
more prominent, indicating stronger interaction of the SN ejecta with the
circumstellar medium. The inspection of archival images from the Panoramic Survey Telescope and Rapid Response System
(Pan-STARRS) survey has revealed a variable source at the SN position with a brightest detection in December 2015 at 
$M_r=-14.66\pm0.17$ mag. This was likely an eruptive phase from the massive progenitor star 
that started from  at least mid-2011, and that produced the circumstellar environment within which the star exploded
as a Type IIn SN.
The overall properties of SN~2018cnf closely resemble those of transients such as SN~2009ip. This 
similarity favours a massive hypergiant, perhaps a luminous blue variable, as progenitor for SN~2018cnf.}

   \maketitle

\section{Introduction}

The very late evolutionary stages of massive stars can be investigated by analysing and modelling the supernova (SN) data (i.e. the spectral 
evolution and the multi-band light curves), or by studying the photometric behaviour of the progenitor star in archival images.
The latter method has become very popular thanks to the public availability of extensive image databases and has allowed the detection
of the quiescent progenitors of a number of core-collapse supernovae \citep[CC SNe,][and references therein]{sma09}, as well as the discovery of extreme photometric variability experienced by very massive stars before exploding as SNe 
\citep[e.g.][]{pas07,pas13,mau13,fra13a,ofe13,ofe14,nyh17}.

%--------------------------------------------------- One column table
   \begin{table*}
      \caption[]{Basic information on spectra collected for SN~2018cnf. The phases are from the $V$-band light curve maximum (see Sect. \ref{sect_photometry}).}
         \label{tab1}
      $$     \begin{tabular}{cccccc}
            \hline
            \noalign{\smallskip}
            Date &  MJD & Phase~(days) & Instrumental configuration & Res~(\AA) & Range~(\AA) \\
\hline                    
  2018-06-17 &  58\,286.2 &  $-7.2$ & GTC+OSIRIS+R1000B+R1000R & 7,8 & 3\,630-10\,350 \\
  2018-06-17 &  58\,286.2 &  $-7.2$ & LT+SPRAT                 & 18  & 4\,000-8\,100 \\
  2018-06-18 &  58\,287.6 &  $-5.8$ & FTN+FLOYDS               & 15  & 3\,500-10\,000 \\
  2018-06-19 &  58\,288.2 &  $-5.2$ & GTC+OSIRIS+R2500R        & 3.2 & 5\,580-7\,680 \\
  2018-06-19 &  58\,288.4 &  $-5.0$ & VLT(UT2)+UVES            & 0.15$^\ast$ & 3\,750-9\,460 \\
  2018-06-20 &  58\,289.2 &  $-4.2$ & GTC+OSIRIS+R1000B+R1000R & 7,8 & 3\,630-10\,350 \\
  2018-06-21 &  58\,290.1 &  $-3.3$ & Copernico+AFOSC+gm4      & 14  & 4\,100-8\,200 \\
  2018-06-24 &  58\,293.2 &  $-0.2$ & GTC+OSIRIS+R1000B+R1000R & 7,8 & 3\,630-10\,350 \\
  2018-06-27 &  58\,296.6 &  $+3.2$ & FTN+FLOYDS               & 15  & 3\,500-10\,000 \\
  2018-06-30 &  58\,299.6 &  $+6.2$ & FTN+FLOYDS               & 15  & 3\,500-10\,000 \\
  2018-07-08 &  58\,307.2 & $+13.8$ & GTC+OSIRIS+R1000B+R1000R & 7,8 & 3\,630-10\,350 \\  
  2018-07-08 &  58\,307.6 & $+14.2$ & FTN+FLOYDS               & 15  & 3\,500-10\,000 \\ 
  2018-07-11 &  58\,310.5 & $+17.1$ & FTN+FLOYDS               & 15  & 3\,500-10\,000 \\  
  2018-07-18 &  58\,317.1 & $+23.7$ & Copernico+AFOSC+VPH7     & 15  & 3\,500-7\,280 \\
  2018-07-22 &  58\,321.2 & $+27.8$ & GTC+OSIRIS+R1000B+R1000R & 7,8 & 3\,630-10\,350 \\
  2018-07-23 &  58\,322.8 & $+29.4$ & FTS+FLOYDS               & 23  & 3\,500-10\,000 \\
  2018-07-24 &  58\,323.8 & $+30.4$ & FTS+FLOYDS               & 23  & 3\,500-10\,000 \\
  2018-08-04 &  58\,334.2 & $+40.8$ & GTC+OSIRIS+R2500R        & 3.2 & 5\,580-7\,680 \\
  2018-08-07 &  58\,337.1 & $+43.7$ & Copernico+AFOSC+VPH7     & 14  & 3\,400-7\,280 \\
  2018-08-09 &  58\,339.6 & $+46.2$ & FTN+FLOYDS               & 15  & 3\,500-10\,000 \\  
  2018-08-10 &  58\,340.1 & $+46.7$ & GTC+OSIRIS+R1000B        & 7   & 3\,630-7\,860 \\
  2018-08-18 &  58\,348.2 & $+54.8$ & GTC+OSIRIS+R1000B+R1000R & 7,8 & 3\,630-10\,350 \\
  2018-08-21 &  58\,350.0 & $+56.6$ & GTC+OSIRIS+R1000B        & 7   & 3\,630-7\,860 \\
  2018-08-25 &  58\,355.2 & $+61.8$ & GTC+OSIRIS+R1000B+R1000R & 7,8 & 3\,630-10\,350 \\ 
  2018-09-01 &  58\,362.5 & $+69.1$ & FTN+FLOYDS               & 15  & 3\,500-10\,000 \\ 
  2018-09-09 &  58\,370.2 & $+76.8$ & NTT+EFOSC2+gm11          & 14  & 3\,340-7\,450 \\
  2018-09-12 &  58\,373.1 & $+79.7$ & Magellan+FIRE            &     & 7\,830-25\,200 \\       
  2018-10-10 &  58\,401.2 &$+107.8$ & NTT+EFOSC2+gm11          & 14  & 3\,340-7\,450 \\ 
        \noalign{\smallskip}
            \hline
         \end{tabular}
$$
\tablefoot{
    GTC = 10.4 m Gran Telescopio Canarias (La Palma, Canary Islands, Spain);
    LT = 2.0 m Liverpool Telesope (La Palma, Canary Islands, Spain);
    FTN = 2.0 m Faulkes Telescope North (Haleakala, Hawaii Islands, USA);
    VLT = 8.2 m Very Large Telescope (ESO-Cerro Paranal, Chile); 
    Copernico = 1.82 m Copernico Telescope (Mt. Ekar, Asiago Observatory, Italy);
    FTS = 2.0m Faulkes Telescope South (Siding Spring, Australia);
    NTT = 3.58 m New Technology Telescope (ESO-La Silla, Chile);
    Baade = 6.5 m Magellan Baade Telescope (Las Campanas Observatory, Chile).\\
$^\ast$ Measured at 6300~\AA.

}
   \end{table*}
%----------------------------------------------------------------- 

The CC SNe that experience major variability in the years before the explosion show unequivocal signatures of interaction with circumstellar material 
(CSM) gathered in the late stages of stellar evolution through to the detection of luminous emission in the X-ray, ultra-violet, and radio 
domains. These CC\ SNe also display a very high optical luminosity, accompanied by a slow-evolving light curve, and prominent narrow to 
intermediate-width (10$^2$-10$^3$ km s$^{-1}$) spectral lines in emission \citep[e.g.][]{are99,fra02,str12,kie12,chapoo12,fox13,delarosa16,tad13,smi17b}. 
The lines are produced in the photo-ionised CSM. When the CSM is H-rich,
the spectrum is that of a type IIn SN \citep{sch90,fil97}, while when the CSM is H-poor and He-rich, the spectrum is that of a Type Ibn 
SN \citep{mat00,pas08a,pas16,hos17}. However, SNe showing somewhat
transitional narrow-lined spectra (named Type Ibn/IIn) have also been discovered
\citep{pas08b,smi12,pas15}. 

Although several progenitor scenarios can be proposed to explain the variability of Type IIn SN observables \citep[see e.g.][]{dwa11,lel13,mor13},
a connection between Type IIn SNe and massive luminous blue variables (LBVs) has been suggested by \citet{kot06} and \citet{gal09}, and a number of observations 
seem to support this claim \citep[][and references therein]{smi17}. In particular, multiple outbursts resembling the long-lasting 19th century 
Giant Eruption of $\eta$ Car have been discovered in a  number of cases before the explosion of a Type IIn SN, supporting the existence of a tight link 
between LBVs and SNe IIn.

A recent case is that of SN~2018cnf, which was discovered at a magnitude $g \approx 17.7$ by the All-Sky Automated Survey for Supernovae \citep[ASAS-SN;][]{bri18} on 2018 June 14.34 UT. This object is also known by the survey name ASASSN-18mr\footnote{The source was also detected by the Asteroid Terrestrial-impact Last Alert System (ATLAS), labelled as ATLAS18vyq.}. The SN is located at $\alpha = 23^h39^m31^s.21$ and
$\delta = -03^\circ08'55''.18$ (J2000.0), in the outskirts of its host galaxy LEDA 196096. \citet{pre18} classify the transient as a Type IIn SN, and noticed that it was spatially coincident with PS15dkt, a source with $w$ = 20.72 mag detected by the Pan-STARRS Survey for Transients \citep[PSST,][]{hub15,cha16} on 2015 December 9. It is important to note that pre-SN outbursts are occasionally detected before the explosion of Type IIn SNe \citep[e.g.][]{ofe14}, but the number of well-monitored objects with pre-SN outbursts is still limited. 
This motivated us to trigger an extensive monitoring campaign.
Spectroscopic and photometric follow-up was initiated soon after the classification announcement \citep{pre18}, making use of the facilities available to wide international collaborations, including the extended-Public ESO Spectroscopic Survey for Transient Objects \citep[ePESSTO,][]{sma15} and the NOT Unbiased Transient Survey \citep[NUTS,][]{mat16}.
We also made use of some public data from ATLAS \citep{ton11,ton18} and the ASAS-SN survey \citep{sha14}. 
The observational campaign continued  for about four months until the object became too faint to be comfortably detected with mid-size telescopes.

\section{Host galaxy and reddening} \label{host}

%----------------------------------------------------------------- 
   \begin{figure*}
   \centering
   \includegraphics[width=17.6cm]{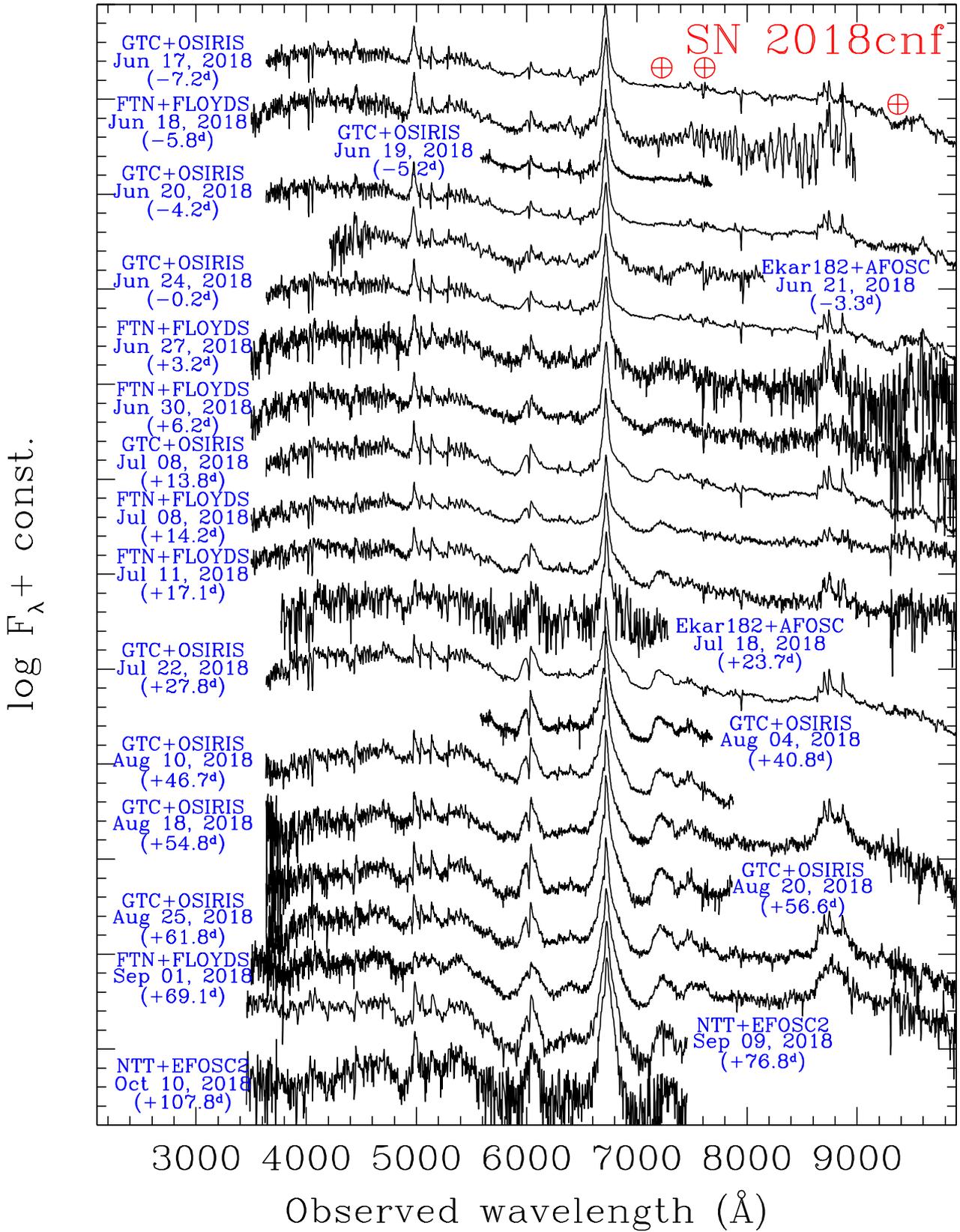}
   \caption{Spectral evolution of SN~2018cnf, spanning approximately 115 day period, from one week before peak to over 100 days after maximum. The phases are from the $V$-band maximum. Only spectra with good signal to noise are shown.}
              \label{fig1}
    \end{figure*}
%----------------------------------------------------------------- 

The host galaxy is an early spiral and the spectrum of its core from \citet{hea09} shows H lines in emission, along with prominent [N II] $\lambda$$\lambda$6548, 6583 lines, and a weak [S II] $\lambda$$\lambda$ 6717, 6731 doublet, 
while [O III] $\lambda$$\lambda$4959, 5007 lines are not detected. Classical absorption features like Ca~II~H$\&$K, Mg I $\lambda$5175, Na ID, and the G-band are also detected.

The redshift of LEDA~196096 is z = 0.02376 $\pm$ 0.00015\footnote{Obtained from the NASA-IPAC Extragalactic Database (NED), {\sl https://ned.ipac.caltech.edu}.}. 
By adopting a recessional velocity corrected for Virgo Infall of 7\,127 $\pm$ 48 km s$^{-1}$ \citep{mou00} and a standard cosmology with H$_0$ = 73 km s$^{-1}$ Mpc$^{-1}$,
we infer a luminosity distance of 99.5 $\pm$ 7.0 Mpc, hence a distance modulus $\mu$ = 34.99 $\pm$ 0.15 mag.
The Milky Way contribution to the total line-of-sight extinction towards  LEDA~196096 is  $E(B-V) = 0.038$ mag, from \citet{sch11} and by adopting the \citet{car89} 
extinction law. Additional host galaxy reddening is inferred from an early, high resolution spectrum of SN~2018cnf. Accounting for this contribution,
we adopt a total reddening $E(B-V) = 0.12\pm0.03$ mag in this paper (see Sect. \ref{sect_highres}).

\section{Spectral evolution} \label{sect_spectroscopy}

Following the classification of SN~2018cnf as a Type IIn event, several spectra were collected spanning a period from one week before maximum to about three months past maximum. We made use of multiple instruments, including the 10.4~m Gran Telescopio Canarias (GTC) equipped with OSIRIS, the 2.0 Liverpool Telescope with SPRAT, the 2.56~m Nordic Optical Telescope (NOT) with ALFOSC, the 1.82~m Asiago Copernico Telescope with AFOSC, the 2.0~m Faulkes North and South Telescopes with the FLOYDS spectrographs of the Las Cumbres Observatory as part of the Global Supernova Project, the 3.58~m New Technology Telescope (NTT) with EFOSC2, and  the 6.5 m Magellan Telescope with FIRE. The spectra were reduced following standard prescriptions in IRAF, including bias, flat-field, and overscan correction. Then we performed optimal extraction of the 1D spectra. Wavelength calibration was done using arc lamp spectra and then was checked using night skylines. Flux calibration was performed with instrumental sensitivity functions obtained from spectro-photometric standards obtained the same night as the SN observation. 
The spectra of the standards were also used to remove telluric absorption bands. Finally, the accuracy of the flux calibration was controlled with coeval photometry (Sect. \ref{sect_photometry}), and a constant factor was applied to the flux in case of a discrepancy. The FIRE NIR spectrum was reduced and telluric-corrected with the procedures described by \citet{hsi19}.
All spectra were finally corrected for a total reddening $E(B-V) = 0.12$ mag, and for a redshift z = 0.023458, as obtained from the position of the narrow Na~I $\lambda\lambda$5890, 5896 (Na~ID) lines of the host galaxy in the high-resolution UVES spectrum of SN~2018cnf (see Sect. \ref{sect_highres}). We note that this redshift value is slightly smaller than that reported by the NED. This can be due to the peripheral location of SN~2018cnf and hence to the host galaxy rotation curve.

The comprehensive information on the dataset is presented in Table \ref{tab1}, and the spectral sequence is shown in Fig. \ref{fig1}. The phases are from the $V$-band maximum light (MJD = 58\,293.4 $\pm$ 5.7, see Sect. \ref{sect_photometry}).

%----------------------------------------------------------------- 
   \begin{figure*}
   \centering
   \includegraphics[width=15cm]{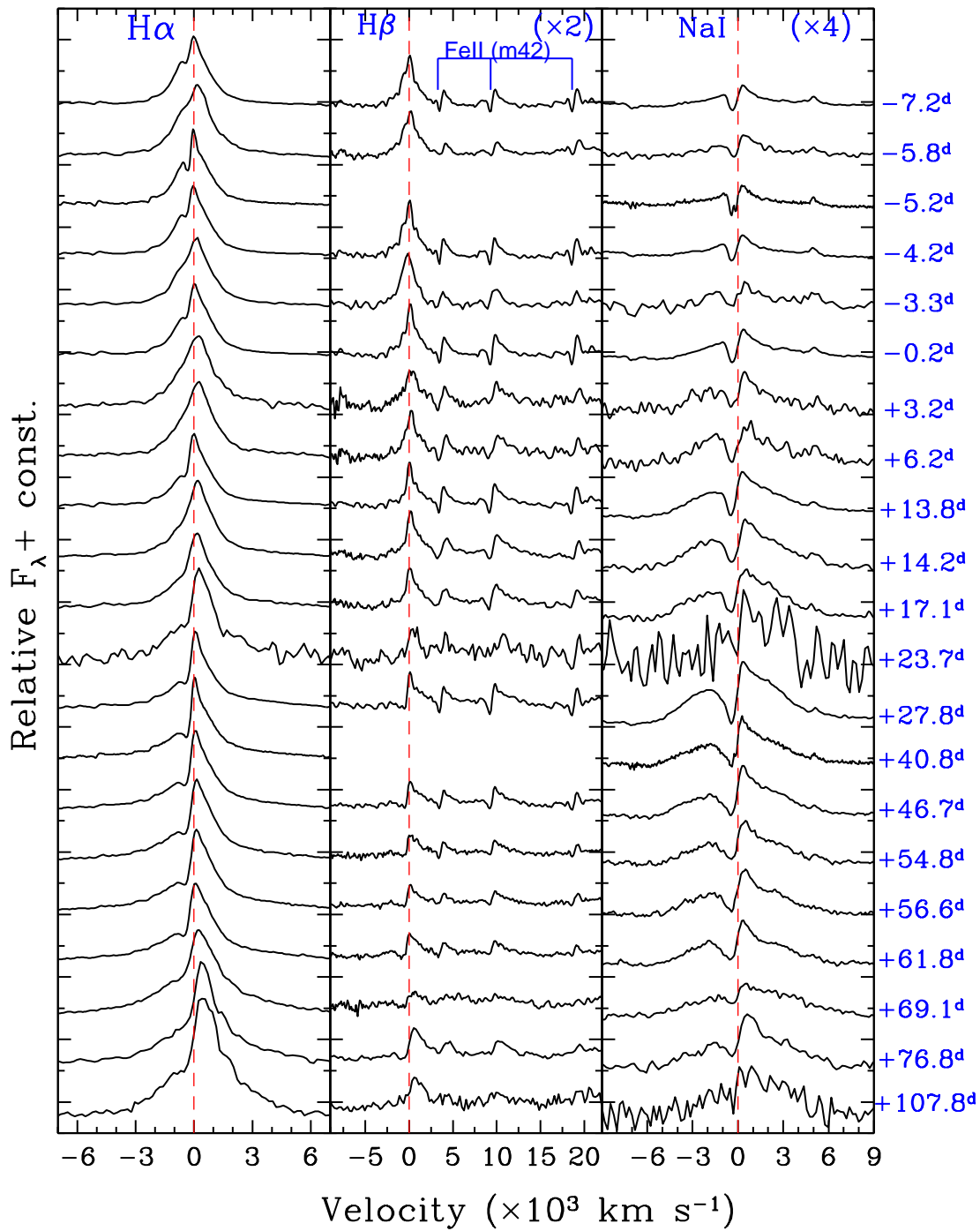}
      \caption{Evolution of profiles of H$\alpha$ (left), H$\beta$ plus Fe~II multiplet 42 region (centre), and Na I doublet plus He~I $\lambda$5876 (right) in best-quality spectra of SN~2018cnf. The red dashed lines mark the zero velocity, corresponding to the rest wavelength of the transition.
              }
         \label{fig2}
   \end{figure*}
%----------------------------------------------------------------

%----------------------------------------------------------------- 
   \begin{figure*}[!t]
   \centering
   {\includegraphics[width=9.5cm,angle=270]{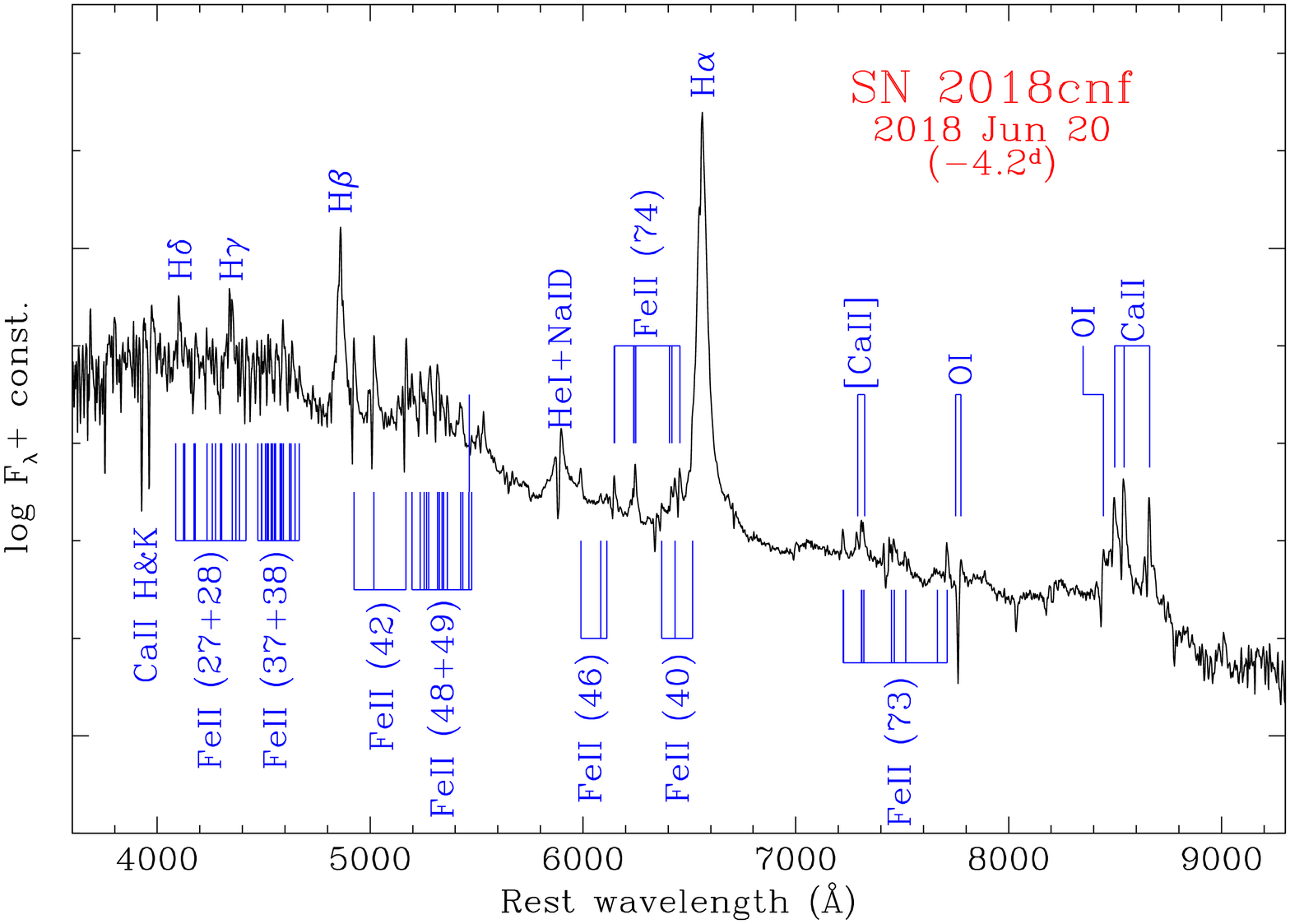}
   \includegraphics[width=9.5cm,angle=270]{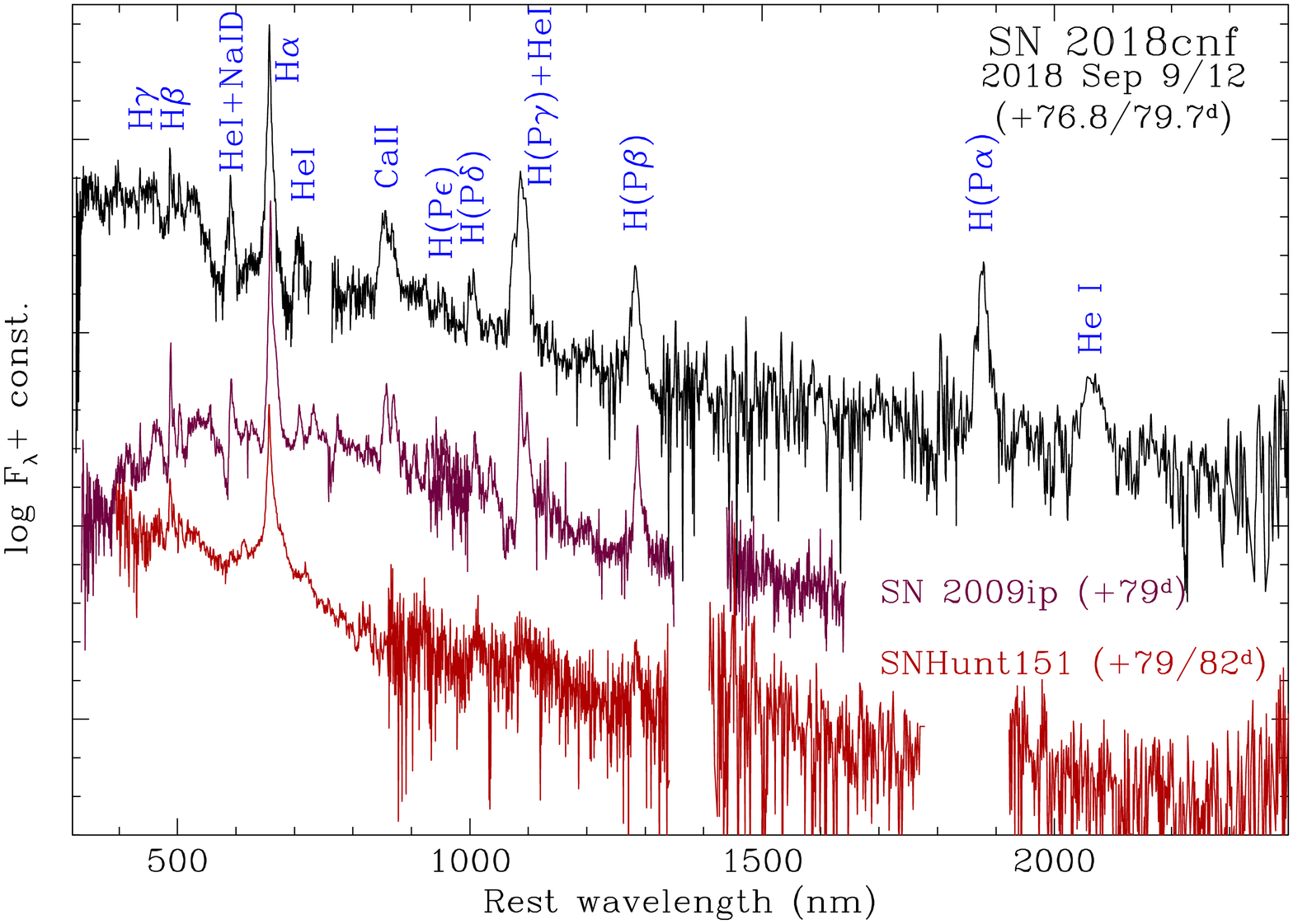}}
      \caption{Top: Line identification in pre-maximum spectrum of SN 2018cnf. A forest of Fe~II lines determines the excess of flux at 
wavelengths shorter than 5400~\AA. Bottom: line identification in the optical plus near-infrared spectra of SN~2018cnf obtained about 80 days after maximum. 
For comparison, also optical plus near-IR spectra of SN~2009ip and SNHunt151 obtained at a similar phase are shown.
}
         \label{fig3}
   \end{figure*}
%-----------------------------------------------------------------

The evolution of the following individual lines is shown in detail in Fig. \ref{fig2}: H$\alpha$, H$\beta$, and the He~I plus Na~I
D feature (right, centre, and left panels, 
respectively). The evolution of the Fe~II lines (triplet 42) is also shown in the central panel of Fig. \ref{fig2}.

Early low-resolution spectra, from one week before to about one week after the maximum light, show a relatively blue continuum (with a black-body temperature of 8\,300 $\pm$ 600 K), 
and superposed strong Balmer and Fe~II lines with  P Cygni profiles with the minimum blue-shifted by about 400 km s$^{-1}$. 
The H lines have two main components: a narrow P Cygni profile with a minimum blue-shifted by $\sim$420 km s$^{-1}$, and an intermediate-width Lorentzian component with a full width at half maximum velocity 
(v$_{FWHM}$) of about 1\,700 km s$^{-1}$.
A similar line profile is also visible in the feature centred at rest wavelength $\sim$5890~\AA.  This feature is very likely Na~I $\lambda\lambda$5890, 5896 (Na~ID). 
While the low resolution spectra show a narrow P Cygni component superposed on the broad base with a minimum blue-shifted by about  370-400~km~s$^{-1}$, the higher resolution of the GTC spectrum
at $-5.2$ d reveals that this narrow P~Cygni has a double-component absorption with one component centred at $\sim$5883\AA~(rest frame), 
and another one centred at $\sim$5888~\AA. These features will be discussed in detail in Sect. \ref{sect_highres}.
Here, we only remark that none of the two absorptions is compatible with He~I. 
A number of P~Cygni metal lines, clearly detected in the early spectra, originate from very strong Fe~II multiplets (Fig. \ref{fig3}, top panel). 
Furthermore, Ca~II is also identified with the H$\&$K  doublet in absorption and the near-infrared (NIR) triplet lines, mostly as narrow emission features.
The O~I features at 7772-7775~\AA~and 8446~\AA~show a strong, narrow component in absorption with a minimum blue-shifted by about 400~km~s$^{-1}$. 
The presence of O~I $\lambda$8446 may be the consequence of Ly$\beta$ pumping due to the Bowen fluorescence mechanism \citep{gra80,fran15}, 
as observed in the Type IIn SN~1995G, for instance \citep{pas02}. However, the simultaneous presence of O~I $\lambda\lambda$7772,7775 favours 
collisional excitation and recombination. 
Alternatively, the presence of these lines can be due to an abundance effect. However, while enhanced abundance of CNO in the stellar envelope 
would require optically thin conditions and accurate verification through synthetic spectral models and/or precise intensity line ratio measurements 
 \citep[in particular using the high-ionisation CNO UV lines; e.g.][]{fra05}, we note that in the early spectra of SN~2018cnf, the line intensity ratio 
H$\beta$ / O~I $\lambda$8446 $\approx 40\pm10$. This is much larger than that observed for SN~1995N, another Type IIn event that showed strong evidence of CNO-burning 
products \citep{fra02}, which is in line with expectations for nearly solar metallicity.

In the following spectra, two to three weeks after maximum, the continuum becomes slightly redder and deviates from a blackbody shape. 
It rather resembles the blue pseudo-continuum produced by blends of Fe II lines, as observed in a number of  ejecta-CSM interacting SNe \citep{smi09}. These spectra
show broader features becoming stronger with time. In particular, the broad 5\,890\AA~feature now has a Gaussian profile, with  v$_{FWHM} \approx$ 5\,350 km s$^{-1}$ at
phase $\sim$15 d, and increases to 5\,850 km s$^{-1}$ at about one month after luminosity peak. 
We also see a broad He~I $\lambda$7065 line, v$_{FWHM} =$ 5\,620 km s$^{-1}$, that does not show a narrow P Cygni component superposed to it. 
For this reason, the broad 5890~\AA~feature at this phase is likely a blend of Na~ID with He~I $\lambda$5876. 
The Balmer lines are still dominated by a narrow to intermediate-width emission\footnote{We remark that, in most spectra, a precise discrimination between
the two components is limited by the modest spectral resolution.} centred at the rest wavelength of the transition;
a narrow absorption component is still detectable. However, a much broader component with the peak slightly shifted towards redder wavelengths 
becomes stronger with time.

The GTC spectrum obtained at phase $+$40.8~d from maximum still shows a broad component for the 5890~\AA~feature with v$_{FWHM} = 5\,380$ km s$^{-1}$.
Superposed to it, we still see a narrow P Cygni feature with the strong minimum centred at $\sim$5883~\AA~and the much weaker redder absorption shoulder centred at $\sim$5888~\AA.
The moderate resolution of this spectrum reveals a more complex profile for the H$\alpha$ line. In fact, we resolve a narrow P Cygni component (v$_{min} = 450$ km s$^{-1}$ 
and v$_{FWHM} = 350$ km s$^{-1}$), an intermediate-width emission with v$_{FWHM} = 1\,850$ km s$^{-1}$, and a broad component centred at 6572~\AA\  with 
v$_{FWHM} = 7\,800$ km s$^{-1}$. The latter component is very likely a direct signature of fast-expanding SN ejecta.

In our spectra obtained from about two months past-maximum, H$\alpha$ shows a narrow Gaussian emission (v$_{FWHM} \approx 700-800$ km s$^{-1}$, with a residual P Cygni absorption) 
superposed on a broader (v$_{FWHM} \approx$ 5\,900 km s$^{-1}$) Gaussian component. The flux contribution of the intermediate-width component is about one order of magnitude 
larger than that of the narrow component. While a narrow P Cygni profile is still clearly visible in the 5890~\AA~feature; this is possibly a blend of He~I $\lambda$5876, but is primarily 
due to Na ID. Furthermore, a much broader base is well fitted by a single Gaussian profile with v$_{FWHM} \approx$ 6\,600 km~s$^{-1}$ , which decreases to about 5\,400 km~s$^{-1}$ one month later. 
The velocity of the broad component is consistent with that inferred from the He~I $\lambda$7065 line. We also note that a broad emission component is now revealed below the
narrow Ca~II NIR triplet, and that both narrow and broad [Ca~II] $\lambda\lambda$7291, 7324 are now identified in these spectra.

In the latest available spectra, phases +76.8 and +107.8 d,  H$\alpha$ can be fitted with two Gaussian components in emission
as the absorption component has almost disappeared. The narrow emission is still quite strong at v$_{FWHM} \approx 700-800$ km s$^{-1}$, while
the broad component has FWHM that decreases from  v$_{FWHM} \sim 4\,800$ to $\sim 4\,500$  km s$^{-1}$ in the two epochs.

   \begin{figure*}
   \centering
   {\includegraphics[width=8.7cm]{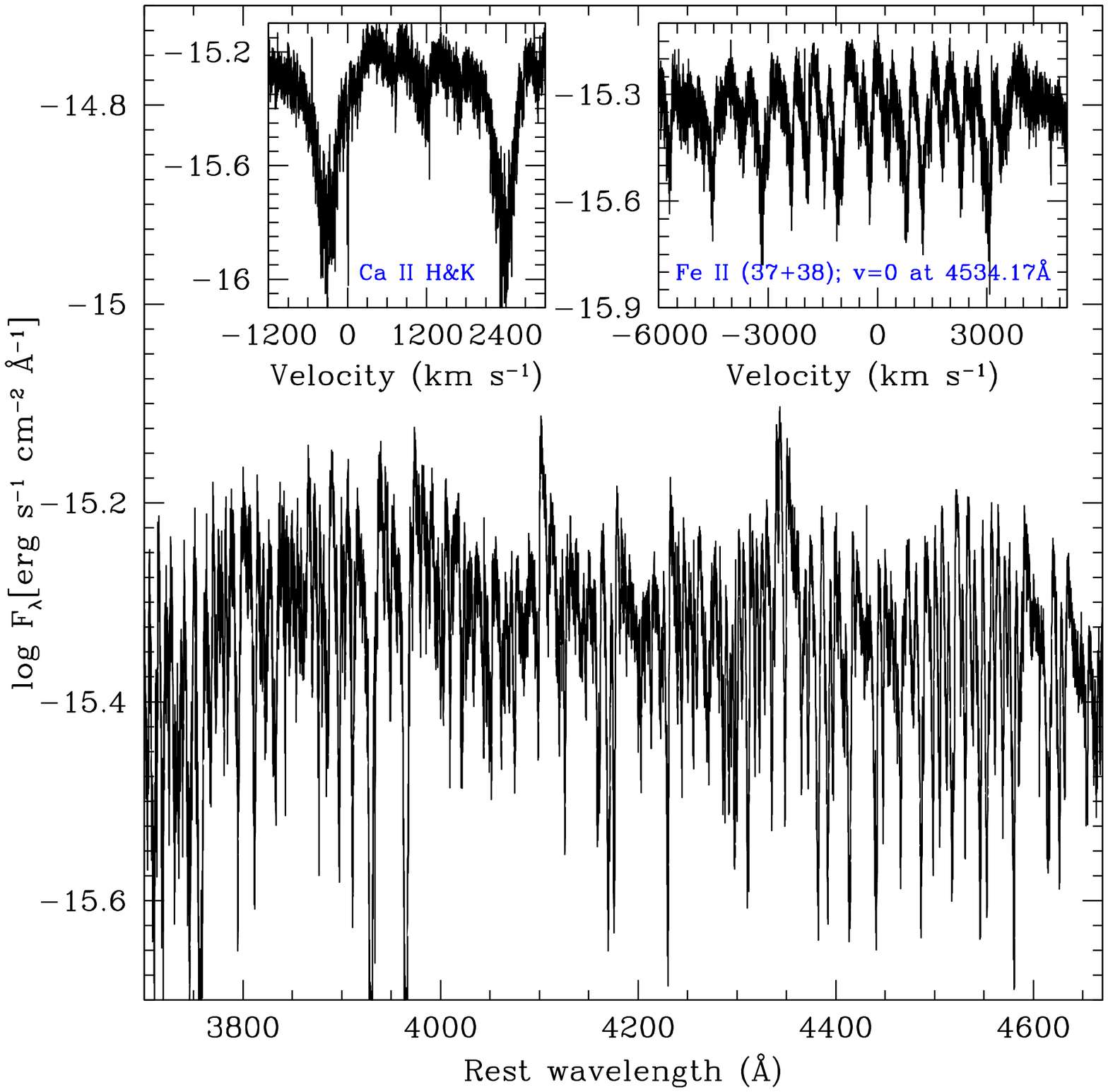}
\includegraphics[width=8.7cm]{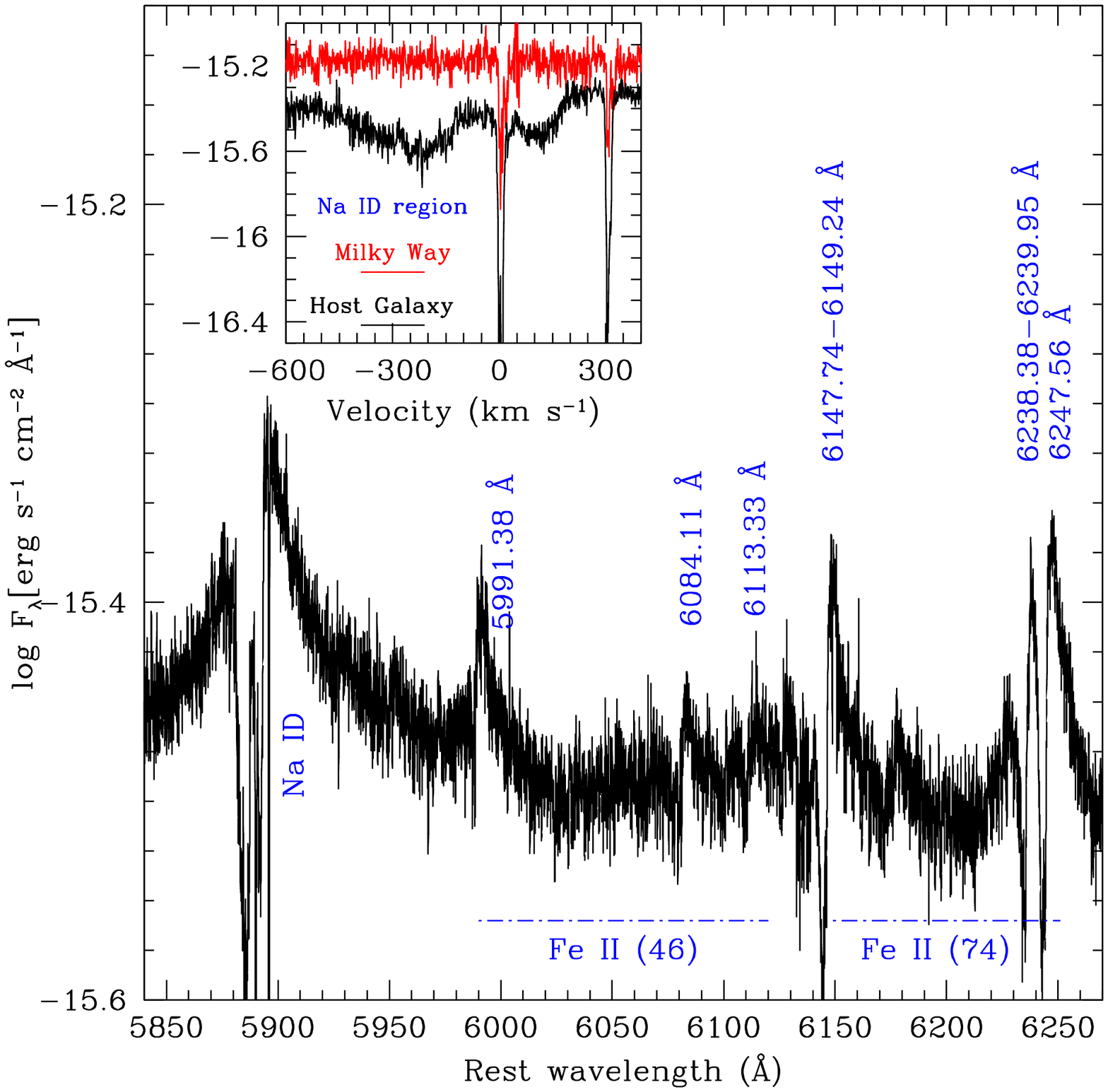}
\includegraphics[width=8.7cm]{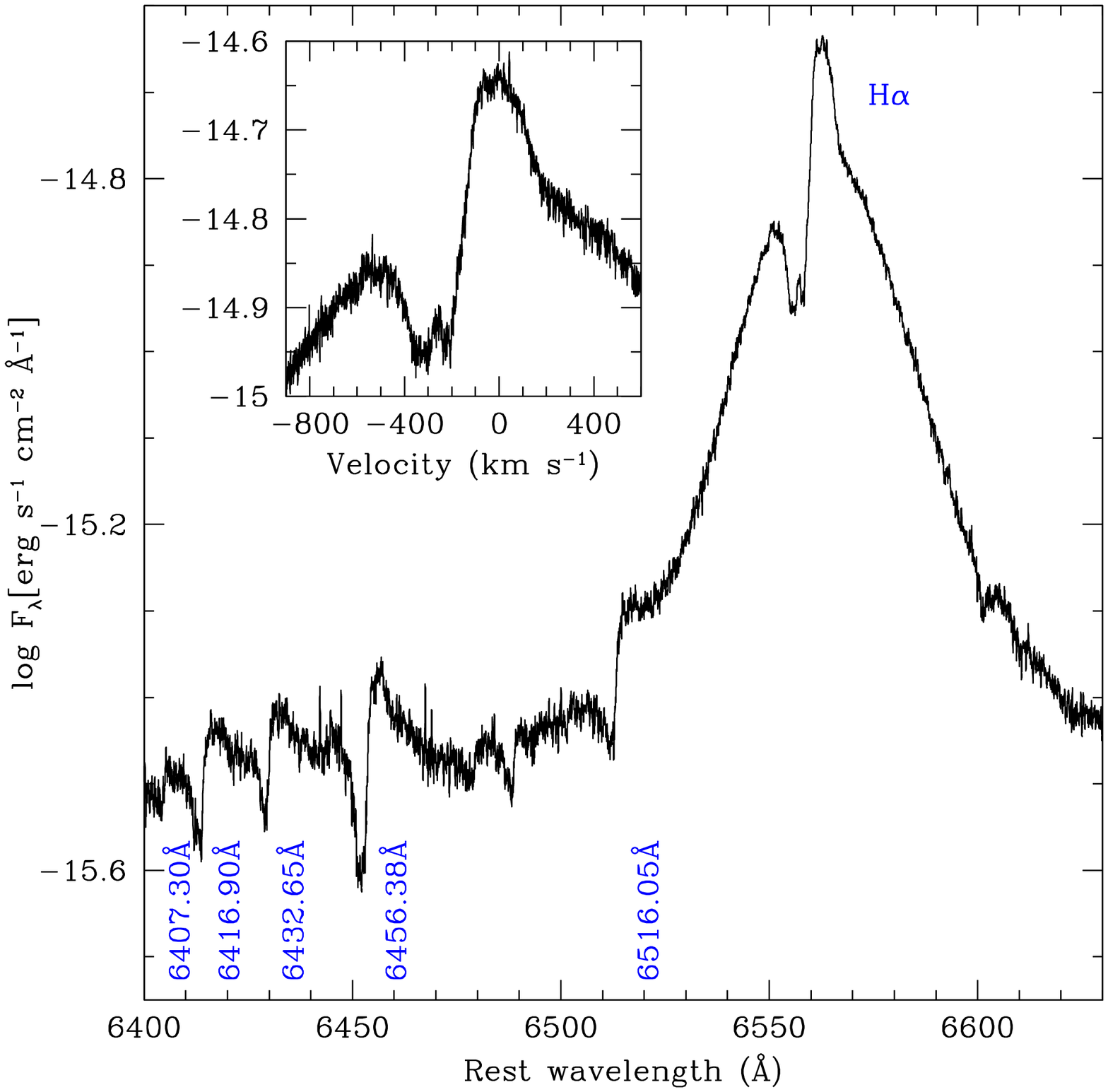}
\includegraphics[width=8.7cm]{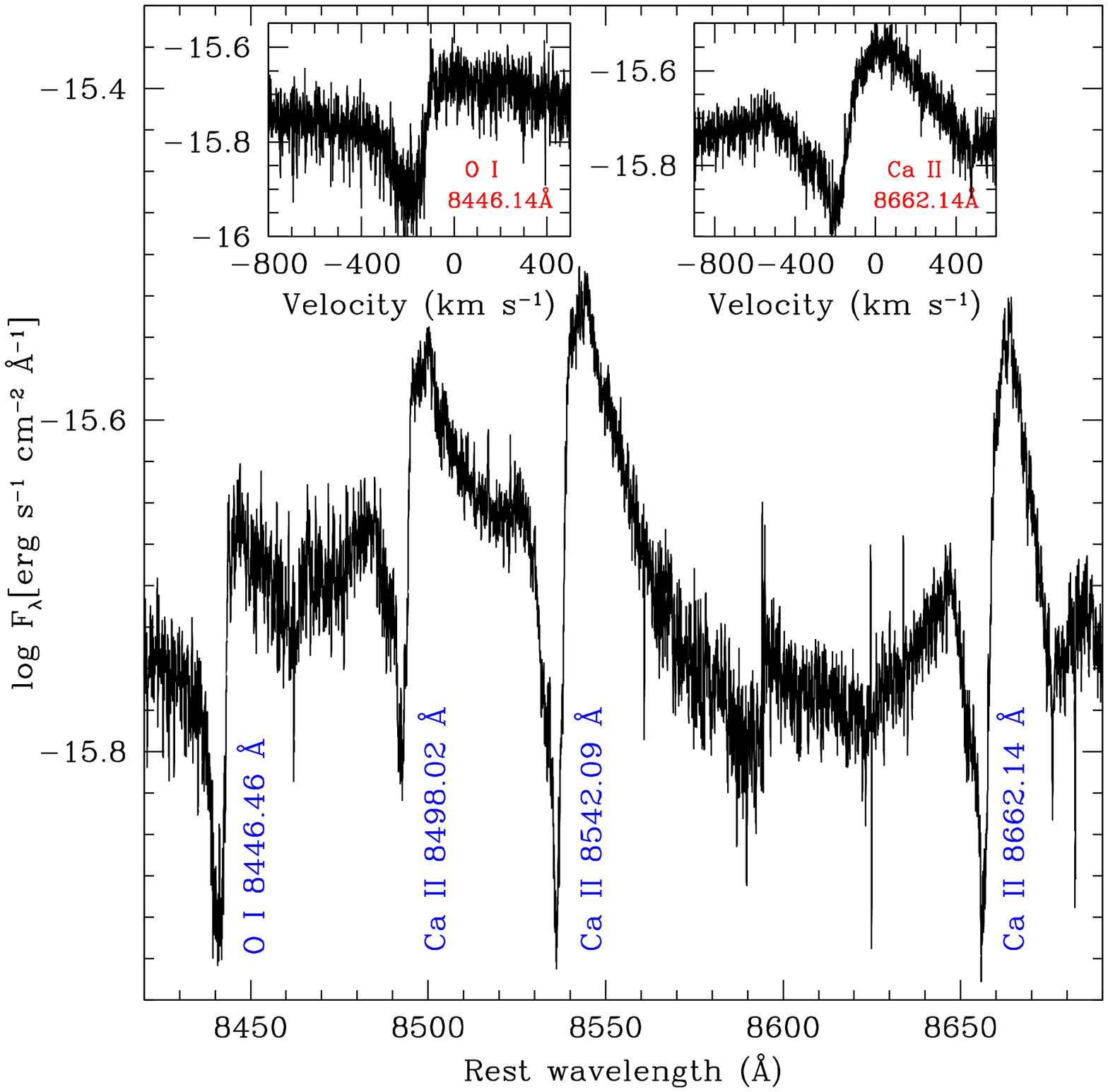}}
   \caption{High resolution UVES spectrum at $-5$ d. Top left: blue region (3700-4670~\AA) dominated by Fe~II lines. The insets show blow-ups of the Ca II H$\&$K region (centred at $\lambda$3933.66) and the Fe~II (multiplets 37 and 38; centred on the $\lambda$4534.17 line). Top right: region between 5840~\AA~and 6270~\AA\  with Na~ID and Fe~II lines (multiplets 46 and 74). The inset shows the narrow Na~ID interstellar doublet centred on the $\lambda$5889.95 line; the upper red line is the Milky Way component; the lower black line one is the host galaxy ones). Bottom left: H$\alpha$ region, including also some Fe~II lines. The blow-up shows a detail of the narrow H$\alpha$ P Cygni. Bottom right: 8420-8690~\AA~region with insets showing a detail of O~I $\lambda$8446.46 (left) and Ca~II $\lambda$8662.14.}
              \label{figUV}
    \end{figure*}

Line identification has been performed in our optical and NIR spectra obtained at around 80 days after the peak (see Fig. \ref{fig3},
bottom). At that phase, along with the optical lines described above, we identified a few H lines of the Paschen series in the NIR spectrum,
as well as the following two prominent He~I features: $\lambda$10830 (blended with H P$\gamma$), and $\lambda$20581 (with v$_{FWHM}
\approx$ 5\,450 km s$^{-1}$), consistent with being due to fast-expanding, He-rich SN ejecta. The relatively broad wings of the
Paschen lines seem to be best reproduced with a lower-velocity ($\approx$3\,500 km s$^{-1}$) Lorentzian profile, which is in agreement with what is observed in the optical spectra. 
Past-maximum (approximately +80 d) optical and NIR spectra of two comparison objects, SN~2009ip \citep{fra13b} and SNHunt151  \citep{eli18}, are also shown; both objects show 
major pre-SN photometric variability signatures (see Sect. \ref{sect_photometry}). Their spectra at that phase are quite different from 
those of SN~2018cnf, both in terms of spectral continuum and line
profiles. This suggests line-of-view effects in an asymmetric gas
distribution or some heterogeneity in the CSM 
geometry and density profile for this SN type. Thus, there are different mass-loss histories and, most likely, progenitor parameters.

\section{High-resolution spectroscopy} \label{sect_highres}

We obtained a high resolution spectrum at the Very Large Telescope (VLT-UT2) with UVES, at phase $= -$5.0 d. The narrow absorption components of the Na~ID doublet visible in the UVES spectrum 
were used to precisely estimate the redshift at the SN location.
Individual spectral regions are shown in Fig. \ref{figUV}.
The blue wavelengths (top left panel) are dominated by a forest of P Cygni lines of Fe~II (from multiplets 26, 27, 28, 37, 38). The velocity, as measured from the minimum of several Fe~II
features, is v$_{Fe}$ = 210 $\pm$ 5 km s$^{-1}$. 
It is interesting to note that the Balmer lines and Ca II H$\&$K are also prominent, the latter with an average velocity v$_{Ca}$ = 270 $\pm$ 6 km s$^{-1}$. However, the red part of H$\beta$ is not covered by the blue arm. 
Additionally, Sr~II (multiplets 1 and 3) lines are  tentatively identified with velocity v$_{Sr}$ = 185 $\pm$ 8 km s$^{-1}$.

%----------------------------------------------------------------- 
       \begin{table*}[!t]
      \caption[]{Near-infrared $JHK$ (Vega mag) photometry of SN~2018cnf.}
         \label{tab2}
      $$     \begin{tabular}{cccccc}
            \hline
            \noalign{\smallskip}
            Date &  MJD & $J$ & $H$ & $K$ & Instrument \\ \hline                    
2018-06-20 & 58\,290.33 &  16.917 (0.156) &  16.802 (0.298) &   $--$            &   SMARTS \\
2018-06-28 & 58\,297.17 &  16.838 (0.050) &  16.673 (0.102) &  16.077 (0.049) &   NOT \\
2018-07-06 & 58\,306.35 &  16.890 (0.175) &  16.749 (0.299) &   $--$            &   SMARTS \\
2018-07-14 & 58\,313.10 &  17.402 (0.044) &  17.233 (0.108) &  16.732 (0.063) &   NOT \\
2018-07-31 & 58\,330.12 &  18.185 (0.130) &  17.834 (0.100) &  17.248 (0.088) &   NOT \\
2018-08-12 & 58\,342.20 &  18.502 (0.074) &  18.270 (0.157) &  17.748 (0.082) &   NTT \\
2018-08-19 & 58\,349.21 &  18.670 (0.126) &  18.427 (0.105) &  17.876 (0.187) &   NTT \\
2018-09-03 & 58\,364.27 &  18.684 (0.153) &  18.475 (0.104) &  17.913 (0.180) &   NTT \\
2018-09-09 & 58\,370.05 &  18.707 (0.140) &  18.562 (0.154) &  17.928 (0.173) &   NTT \\
2018-09-17 & 58\,378.28 &  18.844 (0.151) &  18.679 (0.171) &  18.126 (0.268) &   NTT \\
2018-10-16 & 58\,407.07 &  19.694 (0.127) &  19.561 (0.217) &  18.601 (0.143) &   NOT \\
         \noalign{\smallskip}
            \hline
         \end{tabular}
$$
\tablefoot{SMARTS = 1.3 m SMARTS Telescope + ANDICAM (Cerro Tololo Inter-American Observatory, Chile);
NOT = 2.56 m Nordic Optical Telescope + NOTCam (La Palma, Canary Islands, Spain);
NTT = 3.58 m New Technology Telescope + SOFI (ESO-La Silla, Chile).
}
   \end{table*}
%----------------------------------------------------------------- 

As mentioned in Sect. \ref{sect_spectroscopy}, Na~ID has a rather complex profile. The resolution of the UVES spectrum (top right panel in Fig. \ref{figUV}) is sufficient to allow
us to resolve the different line components. In particular, the two broad absorptions visible in the inset are the two deblended Na~I lines intrinsic to the transient and the velocity, which is inferred from the 
wavelength of the minimun, is v$_{Na}$ = 207 $\pm$ 7 km s$^{-1}$, consistent with that obtained for the Fe~II lines. We also note that the blue wing of the absorption dip 
extends up to about 500 km s$^{-1}$. The inset also shows the narrow Na~ID components due to interstellar medium inside LEDA~196096 (black line), as well as that visible
in the spectrum at z = 0 due to the Milky Way (MW; red line). The equivalent width (EW) of the Na~I $\lambda$5889.95 and $\lambda$5895.92 host components are 0.37 and 0.34 \AA, respectively.
In the same way, we measured the EW of the two Galactic components and obtained 0.24 and 0.18~\AA, respectively. The latter are on average over a factor 1.7 smaller than those of the host galaxy.
As the Galactic reddening is $E(B-V)_{MW}=0.038$ mag \citep{sch11}, we obtained a host galaxy contribution of $E(B-V)_{hg} = 0.065$ mag. 
Another approach for estimating the host galaxy reddening is through the empirical relations between the EW of individual Na~I host galaxy lines and $E(B-V)$ from \citet[][see their equations (1), (2) and (3)]{poz12}. These provide a slightly higher mean reddening, $E(B-V)_{hg}=0.098$~mag. The average of the two methods mentioned above gives $E_{hg}(B-V)=0.082$ mag, hence the total line-of-sight extinction towards SN~2018cnf is
$E_{tot}(B-V)=0.12\pm0.03$ mag.

%----------------------------------------------------------------- 
   \begin{figure}
   \centering
   \includegraphics[width=9cm]{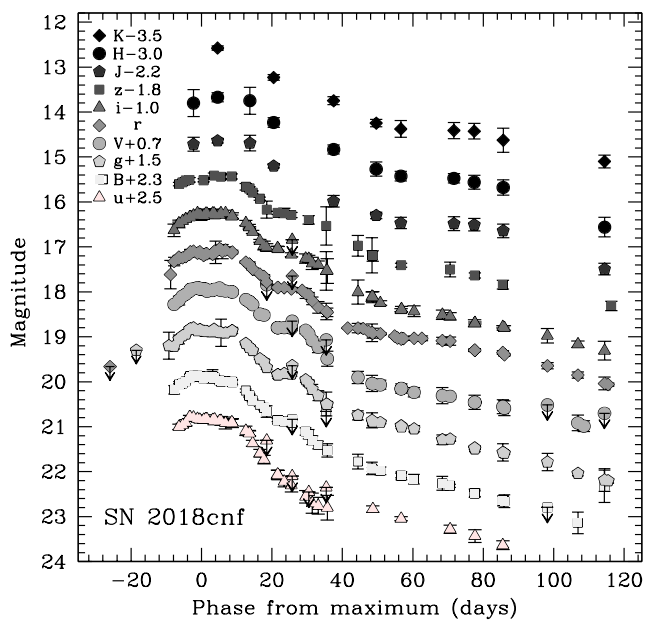}
      \caption{Optical and NIR light curves of SN 2018cnf.}
         \label{lc}
   \end{figure}
%-----------------------------------------------------------------

The bottom left panel of Fig. \ref{figUV} shows the two-component profile of H$\alpha$ with the narrow P Cygni features sitting on a broad Lorentzian base with v$_{FWHM}=1570\pm80$ km s$^{-1}$.
The narrow H$\alpha$ absorption has two dips, one centred at $-$220 km s$^{-1}$, and the second at $-$330 km s$^{-1}$ with a blue wing extending to $\sim$500 km s$^{-1}$. A similar double absorption was first seen in the Type IIn SN 2005gj \citep{tru08}.

The bottom right panel shows the 8400-8700~\AA~region, including O~I $\lambda$8446.46 and the NIR Ca~II triplet.  For both of these species, just as for H$\alpha$ and the Na~ID, 
we find a minimum blue-shifted by 200-220 km s$^{-1}$, and a more structured broader absorption with wings extending to about 500 km s$^{-1}$.
The inspection of the UVES spectrum suggests a pre-SN outflow of gas in expansion towards the observer with a core velocity of 330 km s$^{-1}$ , with a high velocity tail at 500 km s$^{-1}$, 
plus a circum-stellar shell expanding at lower velocity (200-220 km s$^{-1}$). The SN ejecta  likely interact with the faster (330 to 500 km s$^{-1}$) inner CSM, producing the broader Lorentzian component in
emission.

\section{Photometric evolution} \label{sect_photometry}

As the SN was classified about one week before maximum light, we missed the very early SN evolution.  
The multi-filter photometric campaign was triggered soon after the SN classification.
To improve the early-time coverage, we used the information available from the ASAS-SN and ATLAS surveys. Accounting that the earliest 
detection of the SN from ASAS-SN, which is dated 2018 June 14 (MJD = 58\,283.34), the closest but not stringent ASAS-SN pre-SN limit ($g >$ 17.8 mag) is dated 
2018 June 5 (MJD = 58\,274.08), and a relatively deep detection limit (ATLAS-$orange$ $>$ 19.67 mag) was obtained on 2018 May 28 (MJD = 58\,266.61). 
We tentatively assume that SN~2018cnf exploded on MJD = 58\,275.0 $\pm$ 8.4 (i.e. the middle epoch between the deep non-detection on May 28, and the earliest SN detection).

%----------------------------------------------------------------- 
   \begin{figure*}
   \centering
   \includegraphics[width=17.0cm,angle=0]{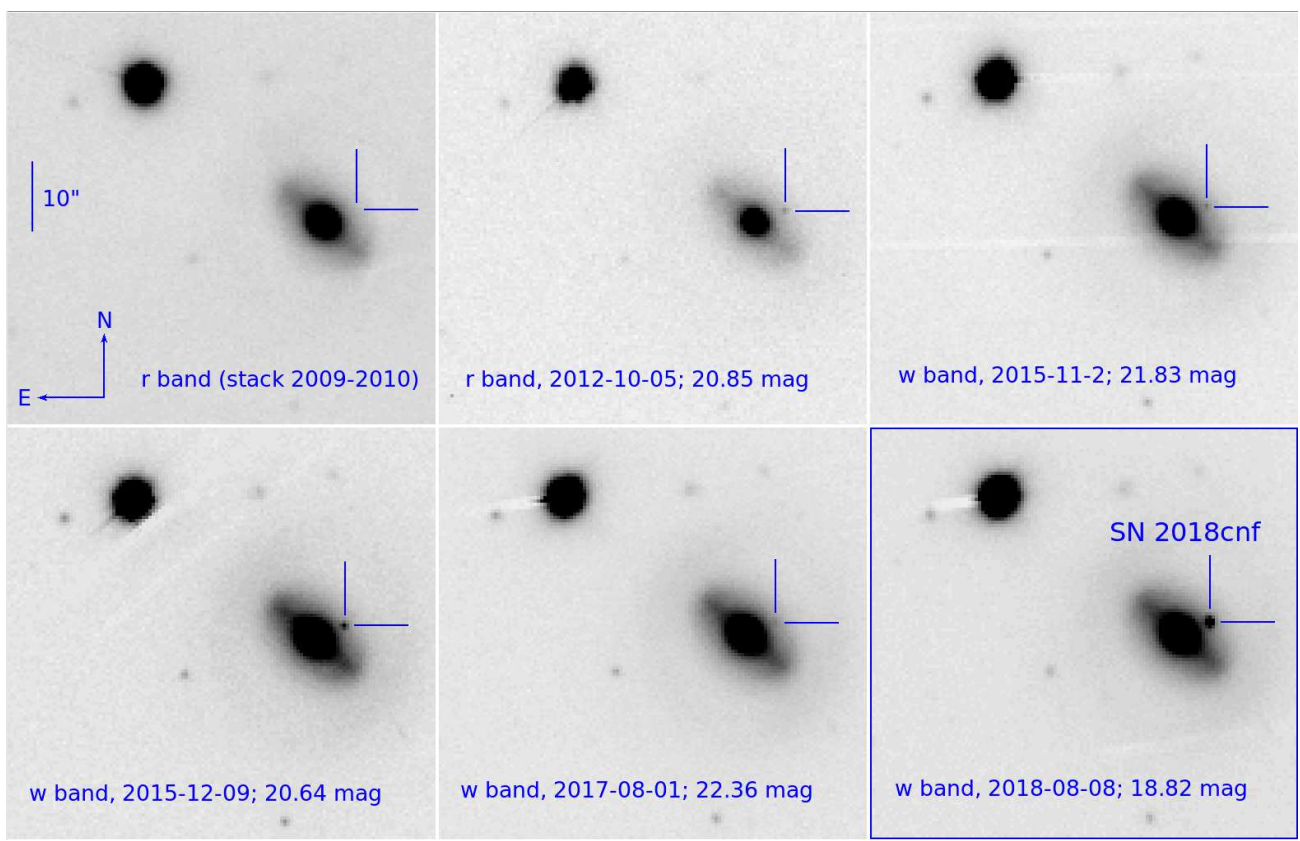}
      \caption{PS1 images of location of SN~2018cnf in Pan-STARRS $r$ and $w$ filters. The top left panel shows a stacked image (limiting magnitude $r > 22.1$ mag) obtained by combining several $r$-band images obtained in 2009-2010 when no sources were visible at the position of SN~2018cnf. Other panels show single-epoch images obtained in October 2012 (top centre), November 2015 (top right), December 2015 (at the time of the brightest outburst; bottom left), August 2017 (bottom centre), and August 2018 (when the SN had already exploded; bottom right). We note the remarkable variability of the progenitor star started at least about six years before the explosion of SN~2018cnf. 
              }
         \label{abs_outburst_figure}
   \end{figure*}
%-----------------------------------------------------------------

The SN photometric measurements were carried out using the SNOoPY pipeline\footnote{SNOoPY is a package for SN photometry using PSF fitting and/or template subtraction developed by E. Cappellaro. A package description can be found at {\it http://sngroup.oapd.inaf.it/snoopy.html}.}, that allowed us to perform the simultaneous PSF-fitting photometry on the SN and a number of stellar sources
of the field. The data in the SDSS bands were then calibrated using
the Sloan catalogue, while $B$ and $V$ photometric data were calibrated using 
a catalogue of SDSS reference stars  and then converted to Johnson-Bessell using the transformation relations of \citet{chr08}. 
Furthermore, NIR photometry was calibrated with reference to the 2MASS catalogue.

The calibrated photometry of SN~2018cnf in the Johnson-Bessell B and $V$ bands (Vega system), the Sloan $u,g,r,i,z,$ and Pan-STARRS $y$ (AB system) bands is reported
in the Appendix (Table \ref{tabA1}), while the NIR ($JHK$) data are in Table  \ref{tab2}.
The final optical  and NIR light curves are shown in Fig. \ref{lc}. 

The object was monitored for only a few days
before maximum, which was reached on MJD = 58\,293.4$\pm$5.7 in the $V$ band. After maximum, the light curves start their declines. The early decline (from two to four weeks
after maximum) is relatively fast, although non-monotonic behaviour is observed with a shoulder detected at 20-30 days past maximum in all bands, but not in the Sloan-$u$ band.
Fluctuations in the post-peak light curve were observed in SN~2009ip \citep{mar15}. They are consistent with the erratic variability displayed during 
the pre-SN phase, indicating a rather complex CSM density profile. 
In this specific case, this is possibly due to the collision of SN ejecta with higher-density CSM (see Sect. \ref{sect_impostor})
expelled in a short-duration, previous mass-loss event.
From about $+$40 d, the SN declines more slowly, with an average decline time of over 2 mag (100d)$^{-1}$ in the bluest bands, while the red optical bands have a slower decline rate of
about 1.6 mag (100d)$^{-1}$. The $r$-band light curve, in particular, has a sort of two-phases late-time decline: from 40 to 70 days the decline rate is
1.1 $\pm$ 0.1  mag (100d)$^{-1}$, and increases to 2.1 $\pm$ 0.1  mag (100d)$^{-1}$ from 70 to 120 days after the $V$-band maximum. The particular behaviour of the $r$-band light curve
 is probably due to dominant contribution of the H$\alpha$ emission whose flux declines quite slowly until day 70.\ It is interesting to note that the H$\alpha$ flux weakens faster at very late phases (Sect. \ref{sect_spectroscopy}).

We have also collected a number of observations in the NIR bands, and the general light curve trend is similar to those observed in the optical bands with 
an initially fast post-peak decline, which is then followed by a slower photometric decline at phases later than about 40 days. Unfortunately, NIR observations are not available 
from about 20 to 40 d, hence we cannot detect post-peak light-curve fluctuations as observed in the optical bands.

%----------------------------------------------------------------- 
   \begin{figure*}
   \centering
   \includegraphics[width=16.0cm,angle=0]{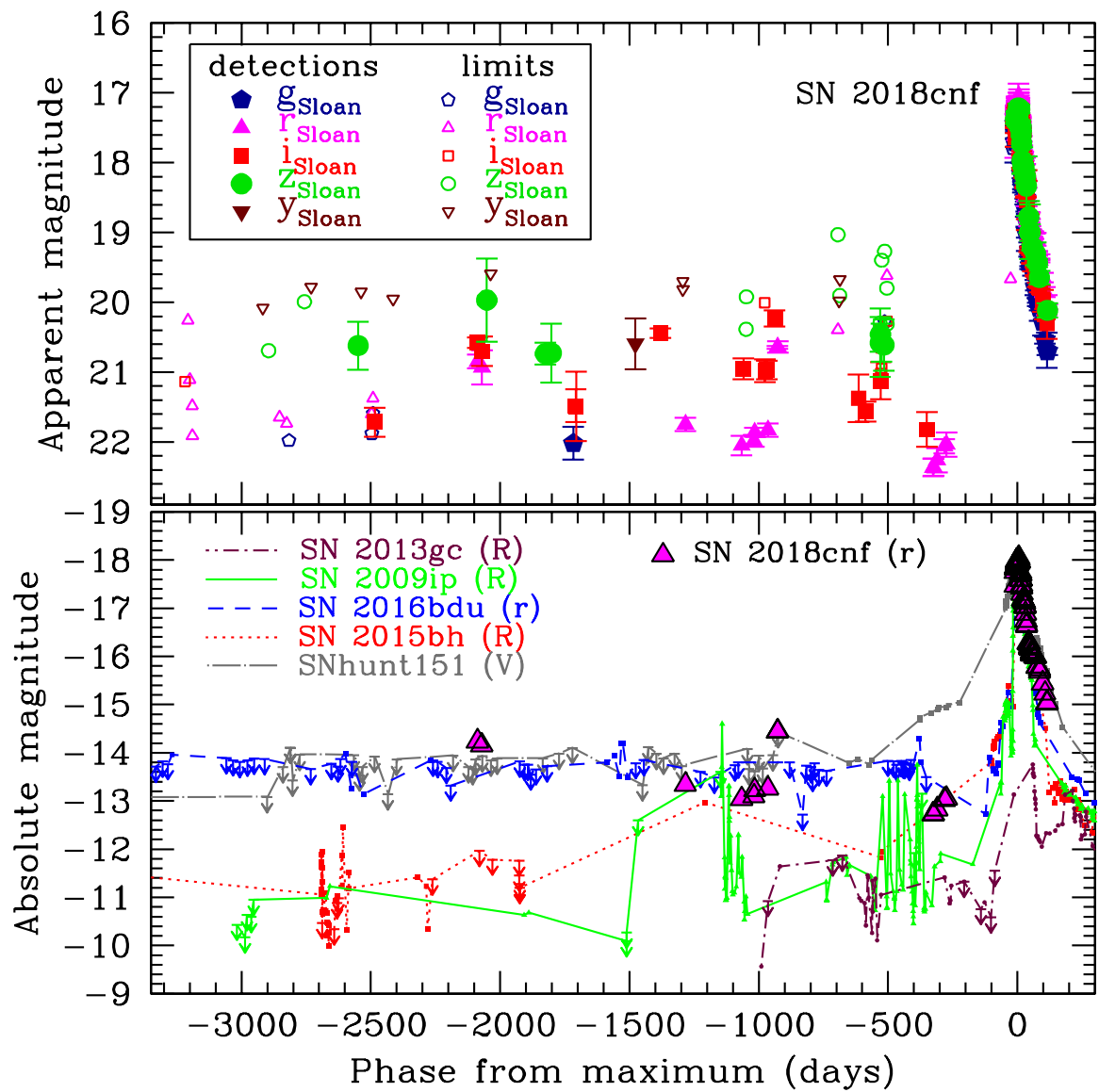}
      \caption{Top: Historical light curve of SN~2018cnf in  $u,g,r,i,z,y$ bands. Filled symbols represent real source detections, open symbols are detection limits. 
Most of the limits reported here, especially those after 2014, were computed from images obtained with relatively poor seeing conditions and/or non-ideal sky transparency.
Occasionally, the number of co-added images also affected the limiting magnitude in the resulting stacked frame. The error-bars on the photometric
points account for uncertainties on the PSF-fit measurements (estimated via artificial star experiments), and those on the nightly zero points and colour terms.
Bottom: Comparison of the $r$-band light curve of SN~2018cnf with those of similar type IIn SNe: SN~2013gc \citep{reg19}, SN~2009ip \citep{pas13,pri13,fra13b,fra15,mar14,gra14,gra17}, SN~2016bdu \citep{pas18}, SN~2015bh \citep{eli16,tho17}, and SNhunt151 \citep{eli18}. For SN~2018cnf, only real detections are shown.
              }
         \label{abs_outburst}
   \end{figure*}
%-----------------------------------------------------------------

In agreement with what is observed in other SNe IIn, the colour evolution of SN~2018cnf is very slow. This agrees with its modest spectral energy distribution (SED; see Sect. \ref{sect_discussion}) evolution.
In particular, the $V-K$ colour shows a slow, almost linear increase from about 0.9 mag soon after the $V$ band maximum to about 1.5 mag three months later.
This, and a lack of evidence for a blue-shift of the spectral emission lines argue against any dust formation in the early months of the SN evolution.
The $B-V$ colour evolution is interesting since it rises from 0.2 to 0.5 mag during the first $\sim 3-4$ weeks after maximum, but then it becomes
slightly bluer ($B-V \approx$ 0.15 mag) at about 40 days after peak. We note that this blueward trend is observed at roughly the same epoch as the post-peak light curve
shoulder. All of this is consistent with an enhanced ejecta-CSM interaction event, as suggested above. 
After the local minimum, the  $B-V$ colour rises again, reaching 0.5 mag at 108 d.

\section{Pre-explosion variability} \label{sect_impostor}

We have conducted an in-depth analysis of pre-SN PS1 images following the brightening detection of 
a stellar source at the position of SN~2018cnf in December 2015 \citep[][and Fig. \ref{abs_outburst_figure}, bottom left panel]{pre18}. 
We inspected a number of PS1 images taken with different filters between August 2009 and September 2017. 
Along with more classical $griz$ images, the SN field was also observed in the Pan-STARRS $y$ and the broad $w$ bands. 
Images obtained from 2009 to mid-2011 showed no source at the position of SN~2018cnf. This was also confirmed by 
combining the best seeing images of that period to construct a deep 2009 to mid-2011 stack frame, which revealed no source
at the SN location down to a limiting magnitude $r \sim 22.1$ mag. 
We thus used the Sloan Digital Sky Survey images obtained on 2009 September 16 as templates to remove the host galaxy contamination 
from the $u, g, r, i, z$ PS1 images at later epochs (from July 2011 to September 2017). Also, pre-SN photometry was obtained with SNOoPY, 
using the PSF-fitting technique on the template-subtracted PS1 images, and the magnitudes were calibrated using reference
stars of the Sloan catalogue. Since good quality templates were not available for PS1 $y-$ and $w$-band images, we performed
PSF-fitting photometry on those images without subtracting the host galaxy contamination. Furthermore, Pan-STARRS $w$ magnitudes were converted to Sloan-$r$
applying a zero-point shift calculated using reference stars from the Sloan catalogue. This did not account for any colour correction and $y$-band magnitudes were obtained applying zero point values computed using reference stars of the Pan-STARRS catalogue.

The earliest marginal detections of a SN precursor were in July and September 2011, although the first robust evidence of
a source at the SN position was in October 2012 (see, Fig. \ref{abs_outburst_figure}), at $r \sim 20.8-20.9$ mag, and $r-i \approx 0.3$ mag. 
We detected the source again in June to July 2013 at $z \approx 20.7$ mag, and in October 2013 at $i \approx 21.5$ mag. A new brightening was detected
in September 2014, reaching $i = 20.44$ mag, then the luminosity of the source settled to $r \sim$ 22 mag and $i \sim 21$ mag
for one year. From late November to late December 2015 (hence over 900 d before the SN light curve
maximum) the object experienced a major brightening, reaching apparent magnitudes $r=20.64$ and $i=20.23$ mag. The former corresponds to a peak absolute magnitude $M_r=-14.66\pm0.17$ mag. This value is rather similar to those observed in
other SN impostors \citep[e.g.][and Fig. \ref{abs_outburst_figure}, bottom left panel]{van00,pas10,smi10,smi11,tar15,tar16a}.

Later, from October 2016 to January 2017 (from about 500 to 600d before the $V$-band peak) the source became fainter with $i > 21$ mag. 
Our latest pre-SN detections showed the progenitor star being even fainter at $i = 21.8$ mag in July 2017 (phase approximately one year before maximum), 
and $r$ possibly varying from $\approx$ 22.3 to 22 mag from August to September 2017 (about 9 months before the SN light curve maximum).
Then, the field became unobservable because of the seasonal gap. It was re-observed only on 2018 May 28 with an initial non-detection 
down to $r > 19.7$ mag, and finally the first SN detection was on 2018 June 14.

A remarkable feature in the SN light curve is the shoulder observed during the post-peak decline (see Sect. \ref{sect_photometry}). Fluctuations
in the post-maximum light curve decline were also detected in SN~2009ip, and were proposed to be correlated with pre-SN stellar activity \citep[e.g.][]{gra14,mar15}.
Although in SN~2018cnf signs of ejecta-CSM interaction are observed from very early phases, a clear flattening is visible
between 2018 July 12 and July 22 (hence approximately on MJD = 58\,311 and 58\,321, respectively), with a middle point on MJD 58\,316 
($\sim$ 41 d after the adopted explosion epoch). This feature is possibly produced by shock-interaction of the SN ejecta with a shell
expelled during a pre-SN stellar outburst.
 By adopting  v$_{ej} = 7\,800$ km s$^{-1}$ (the maximum velocity measured for the broad spectral lines) for the SN ejecta, they reached the innermost CSM layers at 
a distance of 2.8 $\times$ 10$^{10}$ km. In assuming CSM velocities in the range between 200 and 450 km s$^{-1}$ (see Sect. \ref{sect_highres}), this
material was likely ejected from two to about 4.5 years before the collision, which is in agreement with the timescale of the observed pre-SN eruptive phase.

Our analysis on pre-SN images provided numerous detections of a faint, variable source at the SN position starting from at least six years before the SN explosion.
The precursor of SN~2018cnf showed an erratic variability spanning about 2 mag (Fig. \ref{abs_outburst_figure}, top panel). This long-lasting pre-SN variability is similar to 
those observed before the explosion of other peculiar Type IIn SNe \citep[][and references therein]{reg19}, which are usually interpreted
as giant eruptions of massive LBVs \citep[e.g.][]{smi10}.

%----------------------------------------------------------------- 
   \begin{figure}
   \centering
   {\includegraphics[width=9.0cm,angle=0]{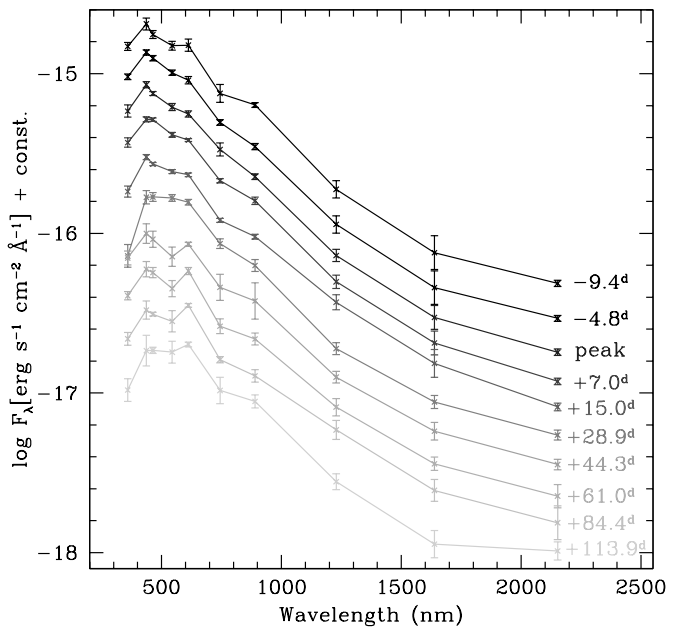}
   \includegraphics[width=9.0cm,angle=0]{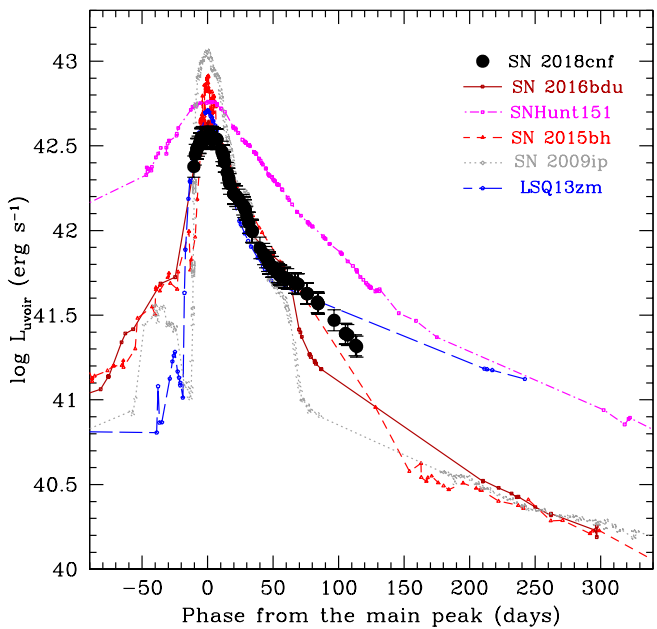}}
      \caption{Top: Evolution of observed SED of SN~2018cnf at some representative epochs. Bottom: Comparison of the quasi-bolometric light curve
of SN~2018cnf with those of a sample of SN~2009ip-like transients, obtained accounting for the contribution in the optical and NIR bands.
In two cases, SN~2009ip and SN~2015bh, we also considered the available UV contribution, which is relatively strong. This explains the more luminous peaks of these
two objects. The phases are with respect to the maximum of the quasi-bolometric light curves. For SN~2018cnf, the peak of the quasi-bolometric light curve is on MJD = 58\,294.2 $\pm$ 0.9.
              }
         \label{sed_bolo}
   \end{figure}
%-----------------------------------------------------------------

\section{Discussion and conclusions} \label{sect_discussion}

The long-term, pre-SN monitoring of the SN~2018cnf host galaxy reveals that the progenitor star experienced a long-lasting eruptive phase before the SN explosion.
This lasted at least a few years, with the source ranging from $M_r \approx -12.8$ to $-14.7$ mag.
During the pre-SN monitoring, a few short-duration outbursts were detected with the most luminous being observed in late 2015,
during which the stellar brightness increased by $\sim$1.2 mag in about one month. This behaviour resembles firstly the strong erratic variability 
we currently see for the impostor SN~2000ch in NGC~3432 \citep{wag04,pas10}, and secondly one of the major brightenings
observed during the pre-SN evolution of SN~2009ip \citep{pas13} or SN~2015bh \citep{ofe16,eli16,tho17}. In both cases, the progenitors are thought
to be LBVs \citep[e.g.][]{smi10,fol11}, although the former experienced a much more dramatic evolution. In fact, SN~2009ip is the 
prototype of a family of transients showing a long-duration variability\footnote{SN~2015bh showed significant variability lasting over two decades \citep{tho17}.}  
resembling the giant eruption of an LBV. 
However, this eruptive phase was followed by an SN-like event a short
time later. This lasted a few months and showed a composite light curve formed by a first
fainter peak at $M_R \geq -15$ mag ('Event A', that was only marginally brighter than the peak magnitude of the former outbursts, but 
lasted approximately two months), and a second much brighter maximum at $M_R \sim -18.5$ mag \citep['Event B'; see figure 5 in][]{pas18}. 
The above double-peaked event is frequently interpreted as the terminal stellar core-collapse signature \citep[][and references therein]{mau13,smi14}.

The recent evolution of SN~2018cnf closely resembles that of several SNe IIn that belong to the SN~2009ip group,
although the spectral energy distribution (SED) of SN~2018cnf has a very slow evolution (Fig. \ref{sed_bolo}, top panel).
Unfortunately, the light curve of SN~2018cnf has not been sampled during the early rise phase to maximum, hence we possibly missed the 
Event A observed in other SN~2009ip-like objects. Another difference with most SN~2009ip-like transients is that the spectrum of SN~2018cnf 
always shows strong signatures of ejecta-CSM interaction and never undergoes a transition towards a more classical broad-lined Type II spectrum ,
as observed in SN~2009ip \citep{pas13,pri13,fra13b,gra14,mar14,fra15,gra17} and similar events, such as
LSQ13zm \citep{tar16b}, SN 2015bh \citep{ofe16,eli16,tho17}, Gaia16cfr \citep{kil18}, and SN~2016bdu \citep{pas18}. The slow spectral evolution is
closely reminiscent of those of iPTF13z \citep{nyh17} and SNHunt151 \citep{eli18}. The comparison of the quasi-bolometric light curves of SN~2018cnf with 
the above Type IIn SN sample is shown in Fig. \ref{sed_bolo} (bottom panel).

As only very massive stars are known to produce similar long-lasting eruptive phases,  a connection with massive LBVs has been proposed
for all of the objects mentioned above. While their progenitors were never directly detected in quiescence, repeated outbursts during which 
the star exceeded $-$14 mag were observed for most of them. 
The physical mechanisms that may produce long-duration giant eruptions, during which LBVs lose a large fraction of their envelope mass
(up to tens of solar masses), have not been clarified yet \citep[e.g.][]{hum94}. Plausible scenarios for these long-duration giant eruptions are major interactions between the massive members 
of a binary system \citep[e.g.][]{smi11b,smi11c}, pulsational pair-instability \citep{cha12,woo17}, or a hybrid scheme of the two \citep[see][]{mar18}. 
While the first mechanism does not necessarily lead to an immediate terminal SN explosion, 
it may explain the three-decade-long eruptive phase of SN~2000ch \citep{pas10}. The second mechanism is promising as it may reproduce the
sequence of events leading to SN~2009ip and its subsequent evolution \citep[e.g.][and references therein]{gra17}.  
Alternatively, mass loss, driven by waves generated in the convective core during the late nuclear burnings, has also been proposed
for moderate-mass (M$_{ZAMS}\sim$20 M$_\odot$) stars and for very massive stars at low-metallicity regime. This
would explain the violent mass ejections observed in some SNe IIn from a decade to a few months prior to core-collapse \citep{qua12,shi14}.

The modern strategy of SN searches and the availability of big data archives have increased the discoveries of pre-SN outbursts, but the
number of events is still of the order of a few tens, and is limited in volume. Future SN searches with big telescopes, in particular
the Large Synoptic Survey Telescope \citep{LSST}, will enormously increase the discovery rates, and will provide us with valuable datasets of
light curves with coverage of many years.

\begin{acknowledgements}

We thank the anonymous referee for insightful comments that helped to improve the paper.

D.A.H, C.M., and G.H. were supported by NSF grant AST-1313484.
S.B., P.C., and S.D. acknowledge Project 11573003 supported by NSFC. This research uses data obtained through the Telescope Access Program (TAP), which has been funded by the National Astronomical Observatories of China, the Chinese Academy of Sciences, and the Special Fund for Astronomy from the Ministry of Finance. 
M.G. is supported by the Polish National Science Centre grant OPUS 2015/17/B/ST9/03167.
T.W. is funded in part by European Research Council grant 320360 and by European Commission grant 730980.
E.Y.H., C.A., and S.K. acknowledge the support provided by the National Science Foundation under Grant No. AST-1613472.
M.F. is supported by a Royal Society - Science Foundation Ireland University Research Fellowship.
M.M.P. acknowledges support from the National Science Foundation under grants AST-1008343 and AST-1613426.
C.T., A.dU.P., D.A.K., and L.I. acknowledge support from the Spanish research project AYA2017-89384-P. C.T. and A.dU.P. 
acknowledge support from funding associated to Ram\'on y Cajal fellowships (RyC-2012-09984 and RyC-2012-09975). 
D.A.K. and L.I. acknowledge support from funding associated to Juan de la Cierva Incorporaci\'on fellowships (IJCI-2015-26153 and IJCI-2016-30940).
G.P and O.R. acknowledge support by the Ministry of Economy, Development, and Tourism’s Millennium Science Initiative through grant IC120009, awarded to The Millennium Institute of Astrophysics, MAS.
L.W. is sponsored, in part, by the Chinese Academy of Sciences (CAS), through a grant to the CAS South America Center for Astronomy (CASSACA) in Santiago, Chile.\\

The NOT data were obtained through the NOT Unbiased Transient Survey (NUTS; {\it http://csp2.lco.cl/not/}), which is supported in part 
by the Instrument Center for Danish Astrophysics (IDA). This work is based, in part, on observations collected at the European Organisation
for Astronomical Research in the Southern Hemisphere, Chile, under ESO programme 0101.D-0202, and as part of PESSTO (the Public ESO Spectroscopic Survey 
for Transient Objects Survey) ESO program 188.D-3003, 191.D-0935, 197.D-1075.
This work also makes use of data from the Las Cumbres Observatory Network as part of the Global Supernova Project;
the Nordic Optical Telescope (NOT), operated on the island of La Palma jointly by Denmark, Finland, Iceland, 
Norway, and Sweden, in the Spanish Observatorio del Roque de los Muchachos of the Instituto de Astrof\'isica de Canarias; 
the 1.82~m Copernico Telescope of INAF-Asiago Observatory; the Gran Telescopio Canarias (GTC), installed in the Spanish 
Observatorio del Roque  de los Muchachos of the Instituto de Astrof\'isica de Canarias, in the Island of La Palma;
the 6.5~m Magellan Telescopes located at Las Campanas Observatory, Chile; and the Liverpool Telescope operated on 
the island of La Palma by Liverpool John Moores University at the Spanish Observatorio del Roque de los Muchachos of the 
Instituto de Astrof\'isica de Canarias with financial support from the UK Science and Technology Facilities Council. 
It is also based in part on observations at Cerro Tololo Inter-American Observatory, National Optical Astronomy Observatory (NOAO), 
which is operated by the Association of Universities for Research in Astronomy (AURA), Inc. under a cooperative agreement with the 
National Science Foundation.\\

ASAS-SN is supported by the Gordon and Betty Moore Foundation through grant GBMF5490 to the Ohio State University and NSF grant AST-1515927. Development of ASAS-SN has  been supported by NSF grant AST-0908816, the Mt. Cuba Astronomical Foundation, the Center for Cosmology and AstroParticle Physics at the Ohio State University, the Chinese Academy of Sciences South America Center for Astronomy (CAS-SACA), the Villum Foundation, and George Skestos.
The Pan-STARRS1 Surveys (PS1) have been made possible through contributions of the Institute for Astronomy, the University of Hawaii, 
the Pan-STARRS Project Office, the Max-Planck Society and its participating institutes, the Max Planck Institute for Astronomy, Heidelberg, and the Max Planck Institute for Extraterrestrial Physics, Garching, The Johns Hopkins University, Durham University, the University 
of Edinburgh, Queen's University Belfast, the Harvard-Smithsonian Center for Astrophysics, the Las Cumbres Observatory Global Telescope 
Network Incorporated, the National Central University of Taiwan, STScI, NASA under Grant No. NNX08AR22G issued through the Planetary 
Science Division of the NASA Science Mission Directorate, the US NSF under Grant No. AST-1238877, the University of Maryland, and Eotvos 
Lorand University (ELTE), the Los Alamos National Laboratory, and the Gordon and Betty Moore Foundation.\\

This research has made use of the NASA-IPAC Extragalactic Database (NED), which is operated by the Jet Propulsion Laboratory, 
California Institute of Technology, under contract with the National Aeronautics and Space Administration.

\end{acknowledgements}

\begin{appendix}

\section{Optical photometry of SN~2018cnf}

\longtab[1]{
\begin{landscape}
\begin{longtable}{ccccccccccc}
\caption{\label{tabA1} Optical photometry of SN~2018cnf: Johnson-Bessell $B,V$ (Vega mag), Sloan $u,g,r,i,z$ and PanSTARRS $y$ (AB mag). }\\
\hline
Date &  MJD & $B$ & $V$ & $u$ & $g$ & $r$ & $i$ & $z$ & $y$ & Instrument \\ 
\hline
\endfirsthead
\caption{continued.}\\
\hline                                
Date &  MJD & $B$ & $V$ & $u$ & $g$ & $r$ & $i$ & $z$ & $y$ & Instrument \\ 
\hline
\endhead
\hline
\endfoot
\endlastfoot
%%%%%%
2009-08-31 &  55\,074.49 &        --      &        --      &        --      &        --      &        --      &  $>$21.14      &        --      &        --      & PS1  \\
2009-09-12 &  55\,086.52 &        --      &        --      &        --      &        --      &  $>$20.26      &        --      &        --      &        --      & PS1  \\
2009-09-18 &  55\,092.50 &        --      &        --      &        --      &        --      &  $>$21.11      &        --      &        --      &        --      & PS1  \\
2009-09-28 &  55\,102.45 &        --      &        --      &        --      &        --      &  $>$21.49      &        --      &        --      &        --      & PS1  \\
2009-09-29 &  55\,103.38 &        --      &        --      &        --      &        --      &  $>$21.91      &        --      &        --      &        --      & PS1  \\
2010-06-27 &  55\,374.61 &        --      &        --      &        --      &        --      &        --      &        --      &        --      & $>$20.08       & PS1  \\
2010-07-20 &  55\,397.61 &        --      &        --      &        --      &        --      &        --      &        --      & $>$20.69       &        --      & PS1  \\
2010-08-31 &  55\,439.51 &        --      &        --      &        --      &        --      &  $>$21.65      &        --      &        --      &        --      & PS1  \\
2010-09-28 &  55\,467.39 &        --      &        --      &        --      &        --      &  $>$21.73      &        --      &        --      &        --      & PS1  \\
2010-10-07 &  55\,476.33 &        --      &        --      &        --      &  $>$21.98      &        --      &        --      &        --      &        --      & PS1  \\
2010-12-06 &  55\,536.22 &        --      &        --      &        --      &        --      &        --      &        --      & $>$19.99       &        --      & PS1  \\
2010-12-31 &  55\,561.26 &        --      &        --      &        --      &        --      &        --      &        --      &        --      & $>$19.78       & PS1  \\
2011-07-01 &  55\,743.62 &        --      &        --      &        --      &        --      &        --      &        --      & 20.620 (0.344) &        --      & PS1  \\
2011-07-12 &  55\,754.60 &        --      &        --      &        --      &        --      &        --      &        --      &        --      & $>$19.85       & PS1  \\
2011-08-23 &  55\,796.52 &        --      &        --      &        --      &  $>$21.88      &  $>$21.60      &        --      &        --      &        --      & PS1  \\
2011-08-28 &  55\,801.52 &        --      &        --      &        --      &  $>$21.59      &  $>$21.38      &        --      &        --      &        --      & PS1  \\
2011-09-03 &  55\,808.55 &        --      &        --      &        --      &        --      &        --      & 21.715 (0.208) &        --      &        --      & PS1  \\
2011-11-13 &  55\,878.34 &        --      &        --      &        --      &        --      &        --      &        --      &        --      & $>$19.95       & PS1  \\
2012-10-05 &  56\,205.44 &        --      &        --      &        --      &        --      & 20.848 (0.099) & 20.575 (0.074) &        --      &        --      & PS1  \\
2012-10-22 &  56\,222.32 &        --      &        --      &        --      &        --      & 20.929 (0.246) & 20.702 (0.211) &        --      &        --      & PS1  \\
2012-11-10 &  56\,241.21 &        --      &        --      &        --      &        --      &        --      &        --      & 19.967 (0.598) &        --      & PS1  \\
2012-11-25 &  56\,256.29 &        --      &        --      &        --      &        --      &        --      &        --      &        --      & $>$19.59       & PS1  \\
2013-06-24 &  56\,467.58 &        --      &        --      &        --      &        --      &        --      &        --      & 20.735 (0.155) &        --      & PS1  \\
2013-07-17 &  56\,490.61 &        --      &        --      &        --      &        --      &        --      &        --      & 20.726 (0.420) &        --      & PS1  \\
2013-10-10 &  56\,575.38 &        --      &        --      &        --      & 22.016 (0.235) &        --      &        --      &        --      &        --      & PS1  \\
2013-10-20 &  56\,585.24 &        --      &        --      &        --      &        --      &        --      & 21.478 (0.233) &        --      &        --      & PS1  \\
2013-10-21 &  56\,586.35 &        --      &        --      &        --      &        --      &        --      & 21.490 (0.496) &        --      &        --      & PS1  \\
2014-06-05 &  56\,813.50 &        --      &        --      &        --      &        --      &        --      &        --      &        --      & $>$18.10       & PS1  \\
2014-06-07 &  56\,815.61 &        --      &        --      &        --      &        --      &        --      &        --      &        --      & 20.593 (0.362) & PS1  \\
2014-09-12 &  56\,912.50 &        --      &        --      &        --      &        --      &        --      & 20.443 (0.067) &        --      &        --      & PS1  \\
2014-12-07 &  56\,998.50 &        --      &        --      &        --      &        --      &        --      &        --      &        --      & $>$19.70       & PS1  \\
2014-12-08 &  56\,999.50 &        --      &        --      &        --      &        --      &        --      &        --      &        --      & $>$19.81       & PS1  \\
2014-12-18 &  57\,009.24 &        --      &        --      &        --      &        --      & 21.745 (0.095) &        --      &        --      &        --      & PS1$^{\ddag}$  \\
2015-07-23 &  57\,226.53 &        --      &        --      &        --      &        --      & 22.048 (0.139) &        --      &        --      &        --      & PS1$^{\ddag}$  \\
2015-07-30 &  57\,233.50 &        --      &        --      &        --      &        --      &        --      & 20.952 (0.148) &        --      &        --      & PS1  \\
2015-08-09 &  57\,243.62 &        --      &        --      &        --      &        --      &        --      &        --      & $>$20.39       &        --      & PS1  \\
2015-08-10 &  57\,244.60 &        --      &        --      &        --      &        --      &        --      &        --      & $>$19.92       &        --      & PS1  \\
2015-09-09 &  57\,274.44 &        --      &        --      &        --      &        --      & 21.984 (0.084) &        --      &        --      &        --      & PS1$^{\ddag}$  \\
2015-09-12 &  57\,277.42 &        --      &        --      &        --      &        --      & 21.861 (0.067) &        --      &        --      &        --      & PS1$^{\ddag}$  \\
2015-10-20 &  57\,315.32 &        --      &        --      &        --      &        --      &        --      & $>$20.01       &        --      &        --      & PS1  \\
2015-10-21 &  57\,316.33 &        --      &        --      &        --      &        --      &        --      & 20.980 (0.162) &        --      &        --      & PS1  \\
2015-10-28 &  57\,323.29 &        --      &        --      &        --      &        --      &        --      & 20.967 (0.131) &        --      &        --      & PS1  \\
2015-11-02 &  57\,328.30 &        --      &        --      &        --      &        --      & 21.828 (0.094) &        --      &        --      &        --      & PS1$^{\ddag}$  \\
2015-11-28 &  57\,354.24 &        --      &        --      &        --      &        --      &        --      & 20.228 (0.117) &        --      &        --      & PS1  \\
2015-12-09 &  57\,365.24 &        --      &        --      &        --      &        --      & 20.644 (0.022) &        --      &        --      &        --      & PS1$^{\ddag}$  \\
2015-12-10 &  57\,366.23 &        --      &        --      &        --      &        --      & 20.639 (0.080) &        --      &        --      &        --      & PS1$^{\ddag}$  \\
2016-07-29 &  57\,598.57 &        --      &        --      &        --      &        --      & $>$20.40       &        --      &        --      &        --      & PS1$^{\ddag}$  \\
2016-07-29 &  57\,598.63 &        --      &        --      &        --      &        --      &        --      &        --      & $>$19.03       &        --      & PS1  \\
2016-08-03 &  57\,603.63 &        --      &        --      &        --      &        --      &        --      &        --      &        --      & $>$19.98       & PS1  \\
2016-08-05 &  57\,605.62 &        --      &        --      &        --      &        --      &        --      &        --      & $>$19.90       & $>$19.67       & PS1  \\
2016-10-17 &  57\,678.37 &        --      &        --      &        --      &        --      &        --      & 21.375 (0.341) &        --      &        --      & PS1  \\
2016-11-13 &  57\,705.35 &        --      &        --      &        --      &        --      &        --      & 21.564 (0.144) &        --      &        --      & PS1  \\
2017-01-08 &  57\,761.22 &        --      &        --      &        --      &        --      &        --      & $>$20.38       & $>$20.30       &        --      & PS1  \\
2017-01-09 &  57\,762.21 &        --      &        --      &        --      &        --      &        --      &        --      & 20.577 (0.487) &        --      & PS1  \\
2017-01-10 &  57\,763.21 &        --      &        --      &        --      &        --      &        --      &        --      & 20.455 (0.248) &        --      & PS1  \\
2017-01-10 &  57\,763.22 &        --      &        --      &        --      &        --      &        --      & $>$21.02       &        --      &        --      & PS1  \\
2017-01-11 &  57\,764.21 &        --      &        --      &        --      &        --      &        --      &        --      & $>$20.40       &        --      & PS1  \\
2017-01-11 &  57\,764.22 &        --      &        --      &        --      &        --      &        --      & 21.121 (0.267) &        --      &        --      & PS1  \\
2017-01-12 &  57\,765.22 &        --      &        --      &        --      &        --      &        --      & $>$20.65       & $>$20.42       &        --      & PS1  \\
2017-01-14 &  57\,767.22 &        --      &        --      &        --      &        --      &        --      & $>$20.94       &        --      &        --      & PS1  \\
2017-01-14 &  57\,767.22 &        --      &        --      &        --      &        --      &        --      &        --      & 20.579 (0.272) &        --      & PS1  \\
2017-01-15 &  57\,768.22 &        --      &        --      &        --      &        --      &        --      &        --      & $>$19.40       &        --      & PS1  \\
2017-01-23 &  57\,776.22 &        --      &        --      &        --      &        --      &        --      &        --      & 20.610 (0.368) &        --      & PS1  \\
2017-01-26 &  57\,779.23 &        --      &        --      &        --      & $>$20.28       &        --      &        --      & $>$19.27       &        --      & PS1  \\
2017-02-04 &  57\,788.22 &        --      &        --      &        --      &        --      & $>$19.62       &        --      & $>$19.80       &        --      & PS1  \\
2017-02-05 &  57\,789.22 &        --      &        --      &        --      &        --      &        --      & $>$20.27       & $>$20.31       &        --      & PS1  \\
2017-07-08 &  57\,942.60 &        --      &        --      &        --      &        --      &        --      & 21.818 (0.249) &        --      &        --      & PS1  \\
2017-08-01 &  57\,966.56 &        --      &        --      &        --      &        --      & 22.363 (0.127) &        --      &        --      &        --      & PS1$^\ddag$  \\
2017-08-03 &  57\,968.55 &        --      &        --      &        --      &        --      & 22.359 (0.123) &        --      &        --      &        --      & PS1$^\ddag$  \\
2017-08-17 &  57\,982.50 &        --      &        --      &        --      &        --      & 22.259 (0.174) &        --      &        --      &        --      & PS1$^\ddag$  \\
2017-09-14 &  58\,010.44 &        --      &        --      &        --      &        --      & 22.061 (0.100) &        --      &        --      &        --      & PS1$^\ddag$  \\
2017-09-22 &  58\,018.40 &        --      &        --      &        --      &        --      & 22.034 (0.174) &        --      &        --      &        --      & PS1  \\
2018-05-28 &  58\,266.61 &        --      &        --      &        --      &        --      & $>$19.67       &        --      &        --      &        --      & ATLAS$^\square$ \\
2018-06-05 &  58\,274.08 &        --      &        --      &        --      & $>$17.80       &        --      &        --      &        --      &        --      & ASAS-SN-5$^\bigstar$\\ 
2018-06-14 &  58\,283.34 &        --      &        --      &        --      & 17.700         &        --      &        --      &        --      &        --      & ASAS-SN-3$^\bigstar$\\  
2018-06-14 &  58\,283.78 &        --      &        --      &        --      &        --      & 17.616 (0.314) &        --      &        --      &        --      & SBIG  \\ %-STL-6303-Brimacombe
2018-06-15 &  58\,284.81 & 17.878 (0.093) & 17.578 (0.068) &        --      & 17.710 (0.061) & 17.326 (0.097) & 17.631 (0.140) &        --      &        --      & fl11 \\
2018-06-17 &  58\,286.18 &        --      &        --      &        --      &        --      & 17.227 (0.030) &        --      &        --      &        --      & GTC \\
2018-06-17 &  58\,286.19 & 17.773 (0.077) & 17.493 (0.099) &        --      & 17.561 (0.098) & 17.256 (0.066) & 17.489 (0.084) &        --      &        --      & fl16 \\
2018-06-17 &  58\,286.20 & 17.762 (0.019) & 17.486 (0.017) & 18.505 (0.061) & 17.546 (0.019) & 17.262 (0.024) & 17.476 (0.025) & 17.402 (0.037) &        --      & LT   \\
2018-06-18 &  58\,287.20 & 17.739 (0.024) & 17.396 (0.016) & 18.467 (0.042) & 17.477 (0.014) & 17.244 (0.025) & 17.427 (0.024) & 17.366 (0.028) &        --      & LT   \\
2018-06-19 &  58\,288.15 & 17.688 (0.029) & 17.353 (0.020) & 18.411 (0.069) & 17.439 (0.014) & 17.218 (0.023) & 17.383 (0.024) & 17.324 (0.055) &        --      & LT   \\
2018-06-19 &  58\,288.21 &        --      &        --      &        --      &        --      & 17.204 (0.029) &        --      &        --      &        --      & GTC \\
2018-06-20 &  58\,289.20 &        --      &        --      &        --      &        --      & 17.166 (0.022) &        --      &        --      &        --      & GTC \\
2018-06-20 &  58\,289.20 & 17.621 (0.019) & 17.287 (0.026) & 18.287 (0.041) & 17.335 (0.015) & 17.144 (0.033) & 17.343 (0.028) & 17.310 (0.051) &        --      & LT   \\
2018-06-20 &  58\,289.43 & 17.573 (0.045) & 17.250 (0.042) &        --      & 17.333 (0.055) & 17.138 (0.075) & 17.340 (0.069) &        --      &        --      & fl03 \\
2018-06-20 &  58\,289.43 & 17.577 (0.039) & 17.247 (0.052) & 18.294 (0.040) & 17.327 (0.040) & 17.110 (0.058) & 17.338 (0.046) &        --      &        --      & fl15$^\Diamond$  \\
2018-06-21 &  58\,290.32 & 17.564 (0.042) & 17.233 (0.050) & 18.293 (0.061) & 17.298 (0.036) & 17.097 (0.035) & 17.300 (0.047) &        --      &        --      & fl15$^\Diamond$  \\
2018-06-22 &  58\,291.43 & 17.565 (0.040) & 17.206 (0.030) & 18.328 (0.052) & 17.366 (0.064) & 17.143 (0.050) & 17.249 (0.044) &        --      &        --      & fl15$^\Diamond$  \\
2018-06-23 &  58\,292.09 & 17.585 (0.027) & 17.218 (0.045) &        --      &        --      &        --      &        --      &        --      &        --      & fl06 \\ 
2018-06-23 &  58\,292.12 &        --      &        --      & 18.342 (0.063) & 17.338 (0.040) & 17.108 (0.029) & 17.290 (0.074) &        --      &        --      & fl06$^\Diamond$  \\ 
2018-06-23 &  58\,292.14 & 17.585 (0.039) & 17.238 (0.040) &        --      &        --      &        --      &        --      &        --      &        --      & fl06 \\ 
2018-06-23 &  58\,292.40 & 17.588 (0.055) & 17.236 (0.028) &        --      & 17.348 (0.045) & 17.149 (0.090) & 17.296 (0.113) &        --      &        --      & fl15 \\ 
2018-06-24 &  58\,293.19 &        --      &        --      &        --      &        --      & 17.190 (0.053) &        --      &        --      &        --      & GTC            \\
2018-06-24 &  58\,293.21 & 17.625 (0.022) & 17.236 (0.026) & 18.347 (0.053) & 17.375 (0.017) & 17.180 (0.014) & 17.309 (0.025) & 17.318 (0.036) &        --      & LT              \\
2018-06-24 &  58\,293.79 & 17.604 (0.041) & 17.242 (0.052) & 18.331 (0.047) & 17.379 (0.032) & 17.155 (0.052) & 17.264 (0.054) &        --      &        --      & fl11$^\Diamond$  \\ 
2018-06-25 &  58\,294.17 & 17.580 (0.046) & 17.286 (0.053) & 18.331 (0.095) & 17.383 (0.029) & 17.153 (0.047) & 17.259 (0.103) &        --      &        --      & fl06$^\Diamond$  \\ 
2018-06-26 &  58\,295.73 & 17.579 (0.149) & 17.267 (0.112) &        --      & 17.385 (0.115) & 17.167 (0.098) & 17.263 (0.107) &        --      &        --      & fl12 \\ 
2018-06-26 &  58\,295.79 & 17.583 (0.037) &     --         & 18.350 (0.044) &        --      &        --      & 17.263 (0.049) &        --      &        --      & fl11$^\Diamond$  \\
2018-06-27 &  58\,296.05 & 17.629 (0.039) & 17.251 (0.082) & 18.348 (0.067) & 17.355 (0.046) & 17.137 (0.031) & 17.277 (0.045) &        --      &        --      & fl06$^\Diamond$  \\
2018-06-27 &  58\,296.17 & 17.604 (0.034) & 17.251 (0.023) & 18.379 (0.075) & 17.331 (0.020) & 17.118 (0.021) & 17.256 (0.020) & 17.214 (0.059) &        --      & LT              \\
2018-06-27 &  58\,296.32 & 17.613 (0.041) & 17.244 (0.042) & 18.337 (0.060) & 17.366 (0.067) & 17.096 (0.025) & 17.298 (0.049) &        --      &        --      & fl15$^\Diamond$  \\
2018-06-27 &  58\,296.62 &        --      &        --      &        --      &        --      & 17.119 (0.250) &        --      &        --      &        --      & ATLAS$^\square$  \\
2018-06-28 &  58\,297.69 & 17.670 (0.032) & 17.214 (0.025) & 18.376 (0.063) & 17.398 (0.022) & 17.041 (0.043) & 17.288 (0.046) &        --      &        --      & fl11$^\Diamond$  \\
2018-06-29 &  58\,298.14 & 17.699 (0.040) & 17.268 (0.019) & 18.367 (0.093) & 17.409 (0.302) & 17.061 (0.025) & 17.248 (0.073) & 17.240 (0.042) &        --      & LT              \\
2018-06-29 &  58\,298.56 &        --      &        --      &        --      &        --      & 17.104 (0.157) &        --      &        --      &        --      & ATLAS$^\square$  \\
2018-06-30 &  58\,299.40 & 17.684 (0.062) & 17.282 (0.078) & 18.419 (0.134) & 17.411 (0.062) & 17.073 (0.077) & 17.246 (0.063) &        --      &        --      & fl03$^\Diamond$  \\
2018-07-02 &  58\,301.17 & 17.699 (0.037) & 17.296 (0.035) & 18.400 (0.072) & 17.369 (0.023) & 17.135 (0.021) & 17.325 (0.033) & 17.233 (0.058) &        --      & LT              \\
2018-07-02 &  58\,301.36 & 17.711 (0.054) & 17.306 (0.052) & 18.423 (0.099) & 17.361 (0.070) & 17.112 (0.038) & 17.311 (0.071) &        --      &        --      & fl15$^\Diamond$  \\
2018-07-06 &  58\,305.18 & 17.896 (0.038) & 17.472 (0.033) & 18.593 (0.105) & 17.649 (0.024) & 17.339 (0.023) & 17.471 (0.057) & 17.462 (0.071) &        --      & LT              \\
2018-07-07 &  58\,306.17 & 18.003 (0.031) & 17.524 (0.044) & 18.644 (0.061) & 17.730 (0.039) & 17.385 (0.049) & 17.557 (0.055) &        --      &        --      & fl16$^\Diamond$  \\
2018-07-07 &  58\,306.21 & 18.017 (0.035) & 17.518 (0.024) & 18.650 (0.058) & 17.739 (0.017) & 17.386 (0.020) & 17.564 (0.017) & 17.502 (0.041) &        --      & LT          \\
2018-07-08 &  58\,307.11 & 18.117 (0.027) & 17.582 (0.026) & 18.867 (0.066) & 17.805 (0.016) & 17.490 (0.019) & 17.667 (0.034) & 17.591 (0.037) &        --      & LT          \\
2018-07-08 &  58\,307.16 &        --      &        --      &        --      &        --      & 17.497 (0.017) &        --      &        --      &        --      & GTC        \\
2018-07-10 &  58\,309.17 & 18.214 (0.039) & 17.797 (0.030) & 19.092 (0.085) & 17.987 (0.018) & 17.604 (0.028) & 17.865 (0.030) & 17.727 (0.036) &        --      & LT          \\
2018-07-11 &  58\,310.34 & 18.279 (0.048) & 17.818 (0.057) & 19.232 (0.095) & 18.055 (0.059) & 17.657 (0.041) & 17.972 (0.098) &        --      &        --      & fl15          \\
2018-07-12 &  58\,311.10 & 18.363 (0.127) & $>$17.14       & $>$18.81       & 18.184 (0.039) & 17.786 (0.033) & 18.012 (0.081) & 17.970 (0.190) &        --      & LT          \\
2018-07-15 &  58\,314.19 & 18.551 (0.033) & 18.096 (0.028) & 19.565 (0.097) & 18.344 (0.035) & 17.901 (0.033) & 18.028 (0.043) & 18.044 (0.035) &        --      & LT          \\
2018-07-15 &  58\,314.34 & 18.557 (0.081) & 18.099 (0.076) & 19.575 (0.105) & 18.348 (0.081) & 17.905 (0.060) & 18.049 (0.056) &        --      &        --      & fl15          \\
2018-07-17 &  58\,316.11 & 18.583 (0.056) & 18.096 (0.047) & 19.721 (0.114) & 18.318 (0.066) & 17.907 (0.064) &        --      & 18.046 (0.104) &        --      & LT          \\
2018-07-19 &  58\,318.13 & 18.614 (0.062) & 18.110 (0.049) & 19.805 (0.148) & 18.330 (0.049) & 17.938 (0.035) & 18.173 (0.043) & 18.092 (0.055) &        --      & LT          \\
2018-07-19 &  58\,318.35 & $>$18.54       & $>$17.95       & $>$19.60       & $>$18.14       & $>$17.65       & $>$17.84       &        --      &        --      & fl03          \\
2018-07-22 &  58\,321.18 &        --      &        --      &        --      &        --      & 17.969 (0.054) &        --      &        --      &        --      & GTC        \\
2018-07-23 &  58\,322.17 & 18.807 (0.084) & 18.158 (0.073) & 20.057 (0.139) & 18.456 (0.062) & 18.038 (0.111) & 18.258 (0.105) &        --      &        --      & fl06          \\
2018-07-24 &  58\,323.05 & 18.866 (0.105) & 18.233 (0.054) & $>$19.95       & 18.527 (0.069) & 18.059 (0.044) & 18.264 (0.078) & 18.198 (0.095) &        --      & LT          \\
2018-07-25 &  58\,324.05 & 18.999 (0.096) & 18.383 (0.082) & 20.202 (0.224) & 18.667 (0.109) & 18.202 (0.087) & 18.368 (0.134) &        --      &        --      & fl16          \\
2018-07-26 &  58\,325.75 & 19.109 (0.084) & 18.532 (0.121) & 20.260 (0.186) & 18.807 (0.110) & 18.345 (0.128) & 18.397 (0.083) &        --      &        --      & fl12          \\
2018-07-29 &  58\,328.03 &  $>$18.36      & $>$18.37       & $>$19.85       & 18.996 (0.265) & 18.437 (0.184) & 18.534 (0.429) & 18.335 (0.424) &        --      & LT          \\
2018-07-29 &  58\,328.31 & 19.228 (0.148) & 18.788 (0.165) & 20.302 (0.273) & 19.005 (0.068) & 18.454 (0.095) & 18.536 (0.274) &        --      &        --      & fl15          \\
2018-08-04 &  58\,334.15 &        --      &        --      &        --      &        --      & 18.804 (0.010) &        --      &        --      &        --      & GTC        \\
2018-08-07 &  58\,337.03 & 19.476 (0.164) & 19.205 (0.134) &        --      & 19.246 (0.102) & 18.808 (0.103) & 19.007 (0.268) & 18.776 (0.228) &        --      & Copernico         \\
2018-08-08 &  58\,338.52 &        --      &        --      &        --      &        --      & 18.816 (0.030) &        --      &        --      &        --      & PS1$^\ddag$  \\
2018-08-10 &  58\,340.05 &        --      &        --      &        --      &        --      & 18.883 (0.029) &        --      &        --      &        --      & GTC        \\
2018-08-11 &  58\,341.02 & 19.630 (0.138) & 19.339 (0.180) &        --      & 19.359 (0.165) & 18.909 (0.195) & 19.104 (0.105) & 18.989 (0.390) &        --      & Copernico         \\
2018-08-11 &  58\,341.23 & 19.653 (0.043) & 19.346 (0.105) & 20.331 (0.072) & 19.394 (0.086) & 18.919 (0.111) & 19.120 (0.123) & 19.001 (0.117) &        --      & NTT        \\
2018-08-13 &  58\,343.39 & 19.683 (0.084) & 19.368 (0.102) &        --      & 19.405 (0.023) & 18.930 (0.022) & 19.248 (0.049) &        --      &        --      & fl03          \\
2018-08-17 &  58\,347.47 &        --      &        --      &        --      &        --      & 18.986 (0.027) &        --      &        --      &        --      & PS1$^\ddag$  \\
2018-08-18 &  58\,348.22 &        --      &        --      &        --      &        --      & 19.025 (0.026) &        --      &        --      &        --      & GTC   \\
2018-08-19 &  58\,349.29 & 19.786 (0.052) & 19.457 (0.062) & 20.547 (0.043) & 19.480 (0.013) & 19.051 (0.038) & 19.407 (0.036) & 19.210 (0.045) &        --      & NTT   \\
2018-08-19 &  58\,349.38 & 19.794 (0.089) & 19.451 (0.103) &        --      & 19.506 (0.030) & 19.031 (0.028) & 19.381 (0.059) &        --      &        --      & fl03     \\
2018-08-19 &  58\,349.99 &        --      &        --      &        --      &        --      & 19.037 (0.036) &        --      &        --      &        --      & GTC   \\
2018-08-22 &  58\,352.97 & 19.868 (0.109) & 19.535 (0.104) &        --      & 19.545 (0.071) & 19.033 (0.050) & 19.433 (0.121) &        --      &        --      & fl16     \\
2018-08-25 &  58\,355.18 &        --      &        --      &        --      &        --      & 19.036 (0.054) &        --      &        --      &        --      & GTC   \\
2018-08-30 &  58\,360.99 & 19.958 (0.154) & 19.609 (0.182) &        --      & 19.789 (0.109) & 19.085 (0.073) & 19.510 (0.102) &        --      &        --      & fl06     \\
2018-09-02 &  58\,363.20 & 20.015 (0.036) & 19.626 (0.066) & 20.782 (0.092) & 19.774 (0.051) & 19.095 (0.092) & 19.546 (0.082) & 19.304 (0.162) &        --      & NTT   \\
2018-09-09 &  58\,370.15 & 20.181 (0.090) & 19.756 (0.158) &        --      & 19.972 (0.030) & 19.296 (0.034) & 19.699 (0.081) &        --      &        --      & fl15     \\
2018-09-09 &  58\,370.23 & 20.190 (0.061) & 19.759 (0.072) & 20.931 (0.142) & 19.993 (0.062) & 19.301 (0.026) & 19.693 (0.025) & 19.435 (0.058) &        --      & NTT   \\
2018-09-17 &  58\,378.19 & 20.350 (0.110) & 19.855 (0.154) & 21.137 (0.098) & 20.072 (0.192) & 19.350 (0.042) & 19.779 (0.025) & 19.643 (0.097) &        --      & NTT   \\
2018-09-17 &  58\,378.61 & 20.360 (0.142) & 19.890 (0.167) &        --      & 20.089 (0.025) & 19.399 (0.028) & 19.804 (0.050) &        --      &        --      & fl11     \\
2018-09-29 &  58\,390.83 &  $>$20.50      & $>$19.82       &       --       & 20.286 (0.192) & 19.645 (0.069) & 19.977 (0.159) &        --      &        --      & fl16     \\
2018-10-08 &  58\,399.48 & 20.838 (0.241) & 20.216 (0.172) &        --      & 20.539 (0.050) & 19.853 (0.069) & 20.168 (0.073) &        --      &        --      & fl12     \\
2018-10-10 &  58\,401.18 &        --      & 20.288 (0.120) &        --      &        --      &        --      &        --      &        --      &        --      & NTT   \\
2018-10-16 &  58\,407.04 & 20.026 (0.360) & $>$ 20.00      &        --      & 20.687 (0.251) & 20.039 (0.141) & 20.309 (0.210) &        --      &        --      & fl15     \\
2018-10-17 &  58\,408.08 &        --      &        --      & 21.999 (0.179) & 20.704 (0.049) & 20.062 (0.036) &        --      &        --      &        --      & NOT   \\
2018-10-17 &  58\,408.96 &        --      &        --      &        --      &        --      &        --      &        --      & 20.115 (0.102) &        --      & NOT   \\
2018-12-28 &  58\,480.74 & $>$ 20.39      & $>$ 20.44      &  $>$ 20.92     &   $>$ 20.27    &  $>$ 19.99     &  $>$ 19.94     &   $>$ 19.37    &        --      & Copernico \\ \hline
\end{longtable}

\tablefoot{PS1 = 1.8 m Pan-STARRS Telescope + GPC1 camera (Haleakala, Hawaii Islands, USA);
ATLAS = 0.5 m ATLAS-1 Telescope + STA-1600 CCD (Haleakala, Hawaii Islands, USA);
ASAS-SN-5 = 0.16 m Payne-Gaposchkin Telescope + FLI ProLine PL230 CCD (Las Cumbres Observatory - South African Astronomical Observatory);
ASAS-SN-3 = 0.16 m Paczy\'nski Telescope +FLI ProLine PL230 CCD ( Las Cumbres Observatory - Cerro Tololo Inter-American Observatory, Chile);
GTC = 10.4 m Gran Telescopio Canarias + OSIRIS (La Palma, Canary Islands, Spain); 
LT = 2.0 m Liverpool Telescope + IO:O (La Palma, Canary Islands, Spain);
fl11 = 1.0 m Telescope (Dome 3) + Sinistro CCD (Las Cumbres Observatory - Siding Spring Observatory, Australia);
fl16 = 1.0 m Telescope (Dome 5) + Sinistro CCD (Las Cumbres Observatory - South African Astronomical Observatory);
fl03 = 1.0 m Telescope (Dome 4) + Sinistro CCD (Las Cumbres Observatory - Cerro Tololo Inter-American Observatory, Chile);
fl15 = 1.0 m Telescope (Dome 2) + Sinistro CCD (Las Cumbres Observatory - Cerro Tololo Inter-American Observatory, Chile);
fl06 = 1.0 m Telescope (Dome 7) + Sinistro CCD (Las Cumbres Observatory - South African Astronomical Observatory);
fl12 = 1.0 m Telescope (Dome 8) + Sinistro CCD (Las Cumbres Observatory - Siding Spring Observatory, Australia);
Copernico = 1.82 m Copernico Telescope + AFOSC (Mt. Ekar, Asiago Observatory, Italy);
NTT = 3.58 m New Technology Telescope + EFOSC2 (ESO-La Silla, Chile);
NOT = 2.56 m Nordic Optical telescope + ALFOSC (La Palma, Canary Islands, Spain).\\
$^\ddag$ PanSTARRS $w$ band converted to Sloan $r$, from TNS;
$^\square$ ATLAS $orange$ band, converted to Sloan $r$, from TNS;
$^\bigstar$ ASAS-SN Sloan $g$, from TNS;
$^\Diamond$ Johnson-Bessell $U$ bands, converted to Sloan $u$.
Las Cumbres Observatory data were taken as part of the Global Supernova Project.}
\end{landscape}
}

\end{appendix}

\end{document}